\documentclass[aps,reprint,superscriptaddress,
nofootinbib,showpacs,amsmath,amssymb]{revtex4-1}

\usepackage{latexsym,graphicx,slashed,bm,dcolumn,color}
\usepackage{blkarray}
\usepackage{multirow}
\usepackage{mathtools}
\usepackage{array,mathtools,amssymb}

\newcommand{\beq}[1]{\begin{equation}\label{#1}}
  \newcommand{\eeq}[1]{\label{#1}\end{equation}}
\newcommand{\be}{\begin{equation}}
  \newcommand{\ee}{\end{equation}}
\newcommand{\bea}{\begin{eqnarray}}
  \newcommand{\eea}{\end{eqnarray}}

\newcommand{\beqn}[1]{\begin{eqnarray}\label{#1}}
  \newcommand{\eeqn}{\end{eqnarray}}
\newcommand{\bd}{\begin{displaymath}}
  \newcommand{\ed}{\end{displaymath}}
\newcommand{\mat}[4]{\left(\begin{array}{cc}{#1}&{#2}\\{#3}&{#4}
    \end{array}\right)}

\def\om{\omega}

\newcommand{\ov}{\overline}
\renewcommand{\to}{\rightarrow}
\renewcommand{\vec}[1]{\mathbf{#1}}
\def\vecB{  \boldsymbol{ B}  } 
\renewcommand{\vec}[1]{ \boldsymbol{#1}}
\newcommand{\vect}[1]{\mbox{\boldmath$#1$}}

\newcommand{\dm}{\varepsilon}

\def\bsig{\mbox{\boldmath $\sigma$} }

\def\cD{{\cal D}}

\def\cP{\mathcal P}

\def\ff{{\rm f}}

\begin{document}

\title{New experimental limits on neutron  -- mirror neutron oscillations  \\ 
in the presence of mirror magnetic field}

\author{Z. Berezhiani}
\affiliation{Dipartimento di Fisica e Chimica, Universit\`a di L'Aquila, 67100 Coppito, L'Aquila, Italy} 
\affiliation{INFN, Laboratori Nazionali del Gran Sasso, 67010 Assergi,  L'Aquila, Italy}

\author{R. Biondi}
\affiliation{Dipartimento di Fisica e Chimica, Universit\`a di L'Aquila, 67100 Coppito, L'Aquila, Italy} 
\affiliation{INFN, Laboratori Nazionali del Gran Sasso, 67010 Assergi,  L'Aquila, Italy}

\author{P. Geltenbort}
\affiliation{Institut Laue-Langevin, 71 Avenue des Martyrs, 38000 Grenoble, France}

\author{I. A. Krasnoshchekova} 
\affiliation{B. P. Konstantinov Petersburg Nuclear Physics Institute of National Research Centre `Kurchatov Institute', 
188300 Gatchina, Leningrad region, Russia} 


\author{V. E. Varlamov} 
\affiliation{B. P. Konstantinov Petersburg Nuclear Physics Institute of National Research Centre `Kurchatov Institute', 
188300 Gatchina, Leningrad region, Russia}

\author{A. V. Vassiljev} 
\affiliation{B. P. Konstantinov Petersburg Nuclear Physics Institute of National Research Centre `Kurchatov Institute', 
188300 Gatchina, Leningrad region, Russia}

\author{O. M. Zherebtsov} 
\affiliation{B. P. Konstantinov Petersburg Nuclear Physics Institute of National Research Centre `Kurchatov Institute', 
188300 Gatchina, Leningrad region, Russia}



\begin{abstract}
Present probes 
do not exclude that the  neutron ($n$) oscillation 
into mirror neutron ($n'$), a sterile state exactly degenerate in mass with the neutron, 
can be a very fast process, in fact faster than the neutron decay itself.
This process is sensitive to the magnetic field. Namely, if the mirror magnetic field $\vec{B}'$ 
exists at the Earth, $n-n'$ oscillation probability can be suppressed or resonantly amplified 
by the applied magnetic field  $\vec{B}$,   
depending on its strength and on the angle $\beta$ between $\vec{B}$ and  $\vec{B}'$. 
We present the results of ultra-cold neutron storage measurements aiming to check the anomalies  
observed in previous experiments which could be a signal for $n-n'$ oscillation in the presence of 
mirror magnetic field $B'\sim 0.1$~G. 
Analyzing the experimental data on neutron loses, we obtain  
a new lower limit on $n-n'$ oscillation time $\tau_{nn'} > 17$~s (95 \% C.L.) for any $B'$ 
between 0.08 and 0.17~G, and $\tau_{nn'}/\sqrt{\cos\beta} > 27$~s (95 \% C.L.) for any $B'$ 
in the interval ($0.06\div0.25$)~G. 
 \end{abstract}

\maketitle

\section{Introduction}

The existence of mirror particles was proposed by Lee and Yang, in the same paper were the 
possibility of parity violation was put forward \cite{LY}, for restoring parity in more extended sense: 
for our particles being left-handed,  {\it Parity} can be interpreted as a discrete 
 symmetry which exchanges them with their right-handed mirror partners.  
Hence the parity, violated in each of ordinary and mirror sectors separately, 
would remain as an exact symmetry between two sectors.  
Kobzarev, Okun and Pomeranchuk \cite{KOP} observed  
that mirror particles cannot have ordinary strong, weak or electromagnetic interactions,  
and so they  must form a hidden parallel world as an exact duplicate of ordinary one. 
This idea was further expanded, with different twists, in many subsequent papers 
 \cite{mirror,Holdom,M-neutrinos,BCV,BB-PRL}. (See reviews \cite{IJMPA};  
 for a historical overview, see also Ref. \cite{Okun}).  

At the basic level, one can consider a theory based on a direct product $G \times G'$ 
of identical gauge factors which can naturally emerge e.g. in the $E_8 \times E'_8$ string theory.  
Ordinary particles belong to the Standard Model $G = SU(3)\times SU(2)\times U(1)$ 
or its grand unified extension, while the gauge interactions 
$G' = SU(3)'\times SU(2)'\times U(1)'$ (or its respective extension) describes mirror particles. 
The total Lagrangian must have a form ${\cal L}_{\rm tot} = {\cal L} + {\cal L}' + {\cal L}_{\rm mix}$ 
where the Lagrangians ${\cal L}$ and ${\cal L}'$, which describe the particle interactions 
respectively in observable and mirror sectors, can be rendered identical by imposing a mirror parity 
$G \leftrightarrow G'$ exchanging ordinary and mirror fermions modulo their chirality. 
Thus, if mirror sector exists,  then all our particles: the electron $e$, proton $p$, neutron $n$, photon $\gamma$, 
neutrinos $\nu$ etc. must have invisible mass degenerate mirror twins: $e'$, $p'$, $n'$, $\gamma'$, $\nu'$ etc. 
which are sterile to our strong and electroweak interactions  but interact with ordinary particles via 
universal gravity.  
 

Mirror matter can be a viable candidate for dark matter \cite{BCV,BB-PRL,IJMPA}. 
The possible interactions between the particles of two sectors (encoded in ${\cal L}_{\rm mix}$), 
as the kinetic mixing  between photon and mirror photon \cite{Holdom} or  interactions 
mediated by heavy messengers coupled to both sectors, as gauge bosons/gauginos of common 
flavor symmetry \cite{PLB98} or common $B-L$ symmetry \cite{Kamyshkov},   
can induce mixing phenomena between ordinary and mirror particles. 
In fact, any neutral particle, elementary or composite, might have a mixing with its mirror twin. 
E.g. the photon kinetic mixing \cite{Holdom} can be searched experimentally 
via the positronium -- mirror positronium oscillation \cite{Gninenko} 
and also via direct detection of dark matter \cite{DAMA}. 
The gauge bosons of common flavor symmetry \cite{PLB98} 
can induce the mixing between the neutral pions and Kaons 
and their mirror partners, also with implications for dark matter direct search \cite{CERN}. 
Three ordinary neutrinos $\nu_{e,\mu\,\tau}$ can oscillate into their (sterile) mirror partners  
$\nu'_{e,\mu\,\tau}$ \cite{M-neutrinos}. The respective mass-mixing terms can emerge 
via  the effective interactions which violate $B-L$ symmetries of both sectors. 
These interactions can be induced via the seesaw mechanism 
by heavy gauge singlet ``right-handed'' neutrinos  \cite{BB-PRL} which 
interact with both ordinary and mirror leptons. 
 On the other hand,  the same $B - L$ non-conserving interactions would induce 
CP violating processes between ordinary and mirror particles and thus generate 
the baryon asymmetries in both sectors \cite{BB-PRL}.  
In this way,  the relation between the dark and observable matter fractions in the Universe, 
 $\Omega'_B /\Omega_B \simeq 5$, can be naturally explained \cite{IJMPA}. 

As it was shown in Refs.~\cite{BB-nn',More}, the present probes do not exclude 
the possibility that oscillation between the neutron $n$ and its mirror twin $n'$ 
is a rather fast process, in fact faster than the  neutron decay. 
The mass mixing, $\dm (\ov{n} n' + \ov{n}' n)$, emerges from
$B$-violating six-fermion effective operators of the type 
$(udd)(u'd'd')/M^5$ involving ordinary $u,d$ and mirror $u',d'$ quarks, 
with $M$ being a cutoff scale related to new physics beyond the Fermi scale. 
As far as the masses of $n$ and $n'$ are exactly equal, 
they must have maximal mixing in vacuum and oscillate
with timescale $\tau_{nn'} = \dm^{-1}\sim (M/10\, {\rm TeV})^5$~s. 
Existing experimental limits or cosmological/astrophysical bounds cannot exclude 
oscillation time $\tau_{nn'} = \tau$ of few seconds \cite{BB-nn'}.\footnote{   
For comparison, the neutron--antineutron oscillation time 
is restricted to be larger than few years by the direct experimental limit  
$\tau_{n\bar n} > 0.9 \times 10^8$~s \cite{Baldo} as well as the indirect limits from the stability of nuclei, 
see Ref. \cite{Phillips} for a recent review.
}
It is of key importance that  in nuclei  $n\to n'$ transition  is forbidden by energy
conservation  and thus nuclear stability bounds give no limitations on $\tau$,   
while for free neutrons $n$--$n'$ oscillation is affected by magnetic fields and coherent interactions
with matter which makes this phenomenon rather elusive~\cite{BB-nn',More}. 
On the other hand, it is striking that $n\to n'$ transitions faster than the neutron decay  
can have far going implications for 
the propagation of ultra-high energy cosmic rays at cosmological distances \cite{UHECR}, 
for the neutrons from solar flares \cite{Nasri}, for primordial nucleosynthesis \cite{BBN} 
and for neutron stars \cite{Anti-DM}.\footnote{In principle, $n-n'$ transition can occur 
not only due to mass mixing term $\dm \ov{n} n' + {\rm h.c.}$ , 
but also due to transitional magnetic moment 
$\delta\mu (F_{\mu\nu} + F'_{\mu\nu}) \ov{n} \sigma^{\mu\nu} n' + {\rm h.c.}$
between $n$ and $n'$ states \cite{Arkady}.  However, we do not discuss this possibility here 
and concentrate only on the mass-mixing effects.} 

The possibility of fast $n-n'$ oscillations can be tested in experiments 
searching for neutron disappearance $n\to n'$ and regeneration $n\to n' \to n$ \cite{BB-nn'} 
as well as via non-linear effects on the neutron spin precession \cite{More}. 
In the ultra-cold neutron (UCN) traps  
 $n\to n'$ conversion can be manifested via  the magnetic field dependence of the neutron loss rates.   
For the UCN flight times between wall collisions $t\sim 0.1$~s, 
the experimental sensitivity can reach $\tau\sim 500$~s~\cite{Pokot} 
(see also Ref.~\cite{Bison} for a recent status of the UCN sources for fundamental physics measurements). 


Several experiments searched for $n-n'$ oscillation with the UCN traps  
\cite{Ban,Serebrov1,Bodek,Serebrov2,Altarev}.   
Following the naive assumption \cite{BB-nn'} that the Earth has no mirror magnetic field,   
these experiments compared the UCN loss rates 
 in \emph{zero} (i.e. small enough) and {\it non-zero} (large enough) magnetic fields. 
In this case the probability of $n$--$n'$ oscillation after a time $t$ depends on the applied 
magnetic field $B$ as 
$P_{B}(t) = \sin^2(\om t)/(\om \tau)^2$, 
$ \om = \frac12 | \mu \vec{B} | = (B/1\,{\rm mG}) \times 4.5\,{\rm s}^{-1}$, 
where $\mu = -6 \cdot 10^{-12}\,$eV/G is the neutron magnetic moment.\footnote{ 
Hereafter we use  natural units, $\hbar = c =1$.
}  
For \emph{small} fields ($B < 1$ mG or so, when $\om t < 1$)   
one has $P_{0} = (t/\tau)^2$, while for {\it  large} fields ($B > 20$~mG or so, when $\om t \gg 1$) 
oscillations are suppressed,  $P_{B} < (1/\tau \om)^2 \ll (t/\tau)^2$.  
In this way, lower bounds on the oscillation time were obtained under the
\emph{no mirror field} hypothesis, the strongest being $\tau > 414$~s  at 90\% CL~\cite{Serebrov1} 
adopted 
by the Particle Data Group~\cite{PDG}.

%
 
However, the above limits become invalid in the presence of mirror matter and/or mirror magnetic field 
~\cite{More}.  
If the Earth possesses mirror magnetic field $B'$, than it would show up as 
uncontrollable background suppressing $n-n'$ oscillation even if the ordinary magnetic field is 
screened in the experiments, i.e. $B=0$. However, if experimental magnetic field is 
tuned as $B\approx B'$, then $n-n'$ oscillation would be resonantly amplified. 
In addition, in this case one could observe the strong dependence of the UCN losses 
on the direction of magnetic field \cite{More}. 

Interestingly, some of the measurements   
 show that the UCN loss rates depend on the magnetic field direction 
 at certain values of magnetic fields, 
in particular the ones  
performed with vertical magnetic fields $B\simeq 0.2$~G reported in Ref. \cite{Serebrov2}.  
The detailed analysis of these experimental data   
indicates towards more that $5\sigma$ deviation from the null hypothesis \cite{Nesti} 
which can be interpreted as a signal for $n-n'$ oscillation in the presence of 
mirror magnetic field $B'\sim 0.1$~G at the Earth.  


A dedicated experiment  \cite{Altarev} tested $n-n'$ oscillation in the presence of mirror magnetic field, 
with  a series of measurements varying the  values of applied (vertical) magnetic field from 0 to 0.125 G.  
Its results, yielding the limit $\tau > 12$~s for any $B'$ less than 0.13 G, 
restrict the parameter space ($\tau, B',\beta$) which can be responsible for the above $5\sigma$ 
anomaly  but do not cover it completely.  

In this paper we report the results of additional measurements aiming to test 
the parameter space related to $5\sigma$ anomaly \cite{Nesti}. 
We essentially repeated the experiment \cite{Serebrov2} with different values of the applied magnetic field.     
New limits on $n-n'$ oscillation time were obtained as a function of mirror magnetic field $B'$ 
which however still leave  some margins for the relevant parameter space. 
The paper is organized as follows. First we discuss $n-n'$ oscillation in the presence 
of mirror magnetic field. Then we describe the experiment and show its results. 
At the end we confront our findings with the results of previous experiments and draw our conclusions.

\section{ Oscillation $n-n'$ in the presence of mirror magnetic fields} 

The hypothesis that the Earth might possess a mirror magnetic field, 
with the strength comparable to the Earth ordinary magnetic field, 
might be a not too exotic possibility. 
The Earth may capture some amount of mirror matter \cite{Anti-DM}
if there exist strong enough interactions between ordinary and mirror particles, 
e.g. due photon--mirror photon kinetic mixing.   
In fact, geophysical data on the Earth mass, moment of inertia, normal mode frequencies etc. 
still allow the presence of dark matter in the Earth with a mass fraction up to $4 \times 10^{-3}$ \cite{Ignatiev}. 
Due to a high temperature in the Earth core, the captured mirror matter can be ionized, at least partially. 
Then the drag of free mirror electrons by the Earth rotation, induced by their Rutherford-like scatterings 
off ordinary  matter, again due to the photon kinetic mixing, 
could give rise to circular mirror currents inducing the mirror magnetic field. 
Such a mechanism of the electron drag was proposed in Ref. \cite{BDT} and applied 
to the generation of the galactic magnetic fields. The dynamo effects could 
additionally enforce the mirror magnetic field at the Earth and also change its configuration, 
so that it could also exhibit significant variations in time \cite{More,Anti-DM}.

When free neutrons propagate in the vacuum but 
ordinary $\vect{B}$ and mirror $\vect{B}'$ magnetic fields are both non-zero and arbitrarily oriented, 
$n$--$n'$ oscillation is described by the Schr\"odinger equation with a $4\times 4$ Hamiltonian: 
\beq{H-osc}
i \frac{d\psi}{dt}= H \psi,  \quad \quad
H  = \mat{  \mu \vect{B} \bsig }{\dm} {\dm} {\mu \vect{B}'  \bsig  } ,
\ee
where $\psi = \big(\psi_n(t),\psi_{n'}(t)\big)$ is the wave function of $n$ and $n'$ in two spin states, 
and $\bsig=(\sigma_x,\sigma_y,\sigma_z)$ are the Pauli matrices.  
The exact calculation of $n-n'$ oscillation probability is given in Ref.~\cite{More}.  
In homogeneous fields  $\vec{B}$ and $\vec{B}'$, 
the probability of $n\to n'$  transition after  
a time $t$ can be conveniently reduced to the formula \cite{Nesti}: 
\beqn{P}
P_{\vec{B} \vec{B}' }(t)  & = & \cP_{BB'}(t)  + \cD_{\vec{B}\vec{B}' }(t)  \nonumber \\
& = & \cP_{BB'}(t)  + D_{BB'}(t) \cos\beta\,,  
\eeqn
where $\beta$ is the angle between the vectors $\vect{B}$ and  $\vect{B}'$ and 
\beqn{PD}
&&  \cP_{BB'}(t) =  
\frac{ \sin^2[(\om-\om')t] }{2  \tau^2 (\om -  \om^\prime)^2 } 
\, + \, \frac{ \sin^2[(\om+\om')t] }{2 \tau^2 (\om + \om')^2}, 
\nonumber  \\ 
&&   D_{BB'}(t)=  
\frac{ \sin^2[(\om-\om')t] }{2  \tau^2 (\om -  \om^\prime)^2 } 
\, - \, \frac{ \sin^2[(\om+\om')t] }{2 \tau^2 (\om + \om')^2} , 
\eeqn
where $\tau = \dm^{-1}$, $ \om = \frac12 | \mu B|$ and  $ \om' = \frac12 | \mu B' |$.   
Hence, for given values $B=\vert \vec{B} \vert$ and $B'=\vert \vec{B}' \vert$ the oscillation amplitude  
(\ref{P}) becomes maximal or minimal respectively when $\vec{B}$ and $\vec{B}'$ are parallel or anti-parallel,  
$\cos\beta = \pm 1$: 
\beqn{maxmin}
&& P_{BB' }^{(+)}(t) ={\cP}_{BB'}(t) + {D}_{BB'}(t) = 
\frac{\sin^2[(\om-\om')t]}{\tau^2 (\om - \om')^2}  , 
\nonumber \\
&& P_{BB' }^{(-)}(t) ={\cP}_{BB'}(t) - {D}_{BB'}(t) = 
\frac{\sin^2[(\om+\om')t]}{\tau^2 (\om + \om')^2}   
\eeqn

 In the experiments one cannot control the mirror magnetic field. 
However, ordinary magnetic field  can be varied and thus the dependence of 
$n-n'$ conversion probability  on $\vect{B}$ can be detected. 
In particular, by reversing the magnetic field direction $\vect{B} \to - \vect{B}$ (i.e. $\beta \to \pi - \beta$) 
the probability (\ref{PD}) becomes $P_{-\vec{B}\vec{B}'} = \cP_{BB'} - D_{BB'} \cos\beta $. 
Therefore, 
$P_{\vec{B}\vec{B}'} - P_{-\vec{B}\vec{B}'} = 2D_{BB'}\cos\beta$ 
is non-zero unless $\cos\beta=0$. 
On the other hand, the sum of probabilities $P_{\vec{B}\vec{B}'} + P_{-\vec{B}\vec{B}'} = \cP_{BB'}$ does 
not depend on the angle $\beta$, and it can be compared with the oscillation probability 
in zero magnetic field, 
$P_{\vec{0}\vec{B}'}=\cP_{0B'}$. 
In particular, if the mirror magnetic field is large enough, 
so that $\omega' t > 1$, which for flight times $t\sim 0.1$~s implies $B' >  1$~mG or so, the 
averaged oscillation probability is $\cP_{0B'} =  \frac12 (\omega'\tau)^{-2}$. 
On the other hand, at $B\approx B'$, when 
$\vert \omega-\omega' \vert \,t < 1$, oscillation probability will resonantly amplify, 
$\cP_{BB'} \approx \frac12 (t/\tau)^2 \gg \cP_{0B'}$. 

When the neutrons are stored in the UCN traps, 
the oscillation probability (\ref{P})  should be averaged over 
the distribution of neutron flight times $t$ between the wall collisions. 
For averaging sinusoidal factors in  (\ref{maxmin})  one can use 
the analytic formula suggested in Ref. \cite{Nesti}: 
\beq{formula}
\langle \sin^2(\omega t) \rangle_t = S(\omega)  = 
\frac12 \left[1 - e^{-2\om^2 \sigma_\ff^2} \cos(2 \om t_\ff) \right], 
 \ee
where $t_\ff = \langle t \rangle$ is the neutron mean free-flight time between wall collisions 
and $ \sigma_\ff^2  = \langle t^2 \rangle - t_\ff^2$. 
For an UCN trap of given geometrical form and sizes these characteristic 
times can be computed via Monte Carlo  simulations, and typically one has $t_\ff, \sigma_\ff \sim 0.1$~s. 
Eq. (\ref{formula})  gives a correct asymptotic behavior for the average probabilities (\ref{maxmin}): 
%
\beq{formula-P}
\ov{P}_{BB' }^{(\pm)} = \frac{S(\om \mp \om')}{\tau^2 (\om \mp \om')^2}. 
 \ee
In the limit $\vert \om - \om' \vert \, t_{\rm f}  \gg 1$ second term in Eq. (\ref{formula}) 
is negligible and one can set $S(\om\pm \om') = 1/2$, which is equivalent to 
averaging of $\sin^2$ factors in (\ref{maxmin}). So we get
%
\beq{PD-simple}
\ov{\cP}_{BB'} =  
\frac{ \om^2 + \om^{\prime 2} }{2  \tau^2 (\om^2 -  \om^{\prime 2} )^2 } , \quad 
\ov{D}_{BB'} = \frac{ \om \om^{\prime} }{ \tau^2 (\om^2 -  \om^{\prime 2} )^2 } \, .
\ee
Explicit form (\ref{formula}) of $S(\om- \om')$ is relevant close to the resonance,  
when $|\omega -\omega'| \, t_{\rm f}  \ll 1$, or   $|B-B'| < 1$~mG. 
Then one gets $S(\om-\om') \approx (\omega-\omega')^2 \langle t^2 \rangle$ and 
thus $\ov{P}_{BB'}^{(+)}  \approx \langle t^2\rangle /2\tau^2$. 
As we show below,  in traps with the homogeneous magnetic field 
the mean probabilities calculated with the analytic approximation (\ref{formula})
agree very well  (about a per cent accuracy) to that obtained via 
Monte-Carlo simulations.


Oscillation $n-n'$ can be tested via magnetic field dependence of UCN losses.  
In the absence of $n-n'$ conversion the number of neutrons $N(t_\ast)$ survived 
after effective storage time $t_\ast$ in the trap from the initial amount should not depend on $\vec{B}$, 
as far as the usual UCN losses during the storage as are the neutron decay, wall absorption or upscattering 
are magnetic field independent in the standard physics framework . 
However, if a neutron between the wall collisions  oscillates into a sterile state $n'$, 
then per each collision the latter can escape from the trap. 
Hence, the amount of survived neutrons in the UCN trap with applied 
magnetic field $\vect{B}$ after a time $t_\ast$ is given by 
$N_{\vec{B}}(t_\ast) =  N(t_\ast)  \exp( -n_\ast  \ov{P}_{\vec{B}\vec{B}'})$,
where $\ov{P}_{\vec{B}\vec{B}'}$ is the average  probability of $n-n'$ conversion between the wall scatterings 
and $n_\ast = n(t_\ast)$ is  the mean number of wall scatterings for the neutrons survived after the 
time $t_\ast$. If the magnetic field  direction  is inverted, $\vect{B} \to -\vect{B}$, 
 then the amount of survived neutrons would become 
$N_{-\vec{B}}(t_\ast) = N(t_\ast)  \exp (- n_\ast  \ov{P}_{-\vec{B}\vec{B}'}) $. 
Since  the common factor $N(t_\ast)$ cancels in the  neutron count ratios,  
asymmetry between $N_{\vec{B}}(t_\ast)$ and $N_{-\vec{B}}(t_\ast)$, 
\beq{AB}
A_{\vec{B}}(t_\ast) = \frac{N_{-\vec{B}}(t_\ast) -
  N_{\vec{B}}(t_\ast)} {N_{-\vec{B}}(t_\ast)+N_{\vec{B}}(t_\ast)} \, , 
\ee 
%
should directly trace the difference 
$\ov{P}_{\vec{B}\vec{B}'} - \ov{P}_{-\vec{B}\vec{B}'}=\ov{\cD}_{\vec{B}\vec{B}'}$ \cite{More}. 
Assuming $n_\ast \ov{\cD}_{\vec{B}\vec{B}'} \ll 1$, we get
\beq{DB}
A_{\vec{B}}(t_\ast)/n_\ast  = \ov{\cD}_{\vec{B}\vec{B}'}=  \ov{D}_{BB'}  \cos\!\beta \, .
\ee 
%
%
On the other hand,  one can compare the average 
$N_B(t_\ast) = \frac12 \big[N_{\vec{B}}(t_\ast) + N_{-\vec{B}}(t_\ast)\big]$ 
with the counts $N_0(t_\ast)$ acquired 
under zero magnetic field: 
\beq{EB}
E_B(t_\ast) = \frac{N_0(t_\ast) - N_B(t_\ast)}{N_{0}(t_\ast) +N_{B}(t_\ast)}  \, . 
\ee
This value measures the difference between the probabilities in zero and non-zero magnetic fields. 
Since $\ov{P}_{\vec{B}\vec{B}'} + \ov{P}_{-\vec{B}\vec{B}'} = 2\ov{\cP}_{BB'} $, we have 
 %
\beq{DeltaB}
E_B(t_\ast)/n_\ast =\ov{\cP}_{BB'} - \ov{\cP}_{0B'} = \ov{\Delta}_{BB'}\, , 
\ee
which should not depend on the magnetic orientation but 
only on its modulus $B = \vert \vec{B} \vert$, an it should be resonantly amplified 
if $B\approx B'$.  
Therefore, measuring $E_B$ at different values of $B$, one can obtain direct limits 
on $n-n'$ oscillation time $\tau$, while by measuring $A_{\vec{B}}$ one in fact 
measures the value $\tau_\beta = \tau/\sqrt{\vert \cos\beta \vert}$, i.e. 
is the oscillation time corrected for the unknown 
angle $\beta$ between ordinary and mirror magnetic fields  $\vect{B}$ and $\vect{B}'$. 
Once again, in ideal conditions these measurements should have no systematic 
uncertainties: measuring the neutron counts in different magnetic configurations but otherwise 
in the same experimental conditions, the effects of the regular neutron loses should cancel 
in the count ratios $A_{\vec{B}}$ and $E_B$, and scanning over different test values of applied magnetic 
field $\vec{B}$ with appropriate statistics, one can obtain 
pretty stringent limits on $\tau$ and $\tau_\beta$ as a function of mirror magnetic field $B'$.


\section{Experiment and measurements} 

\begin{figure}[t]
 \begin{center}
\includegraphics[width=8cm]{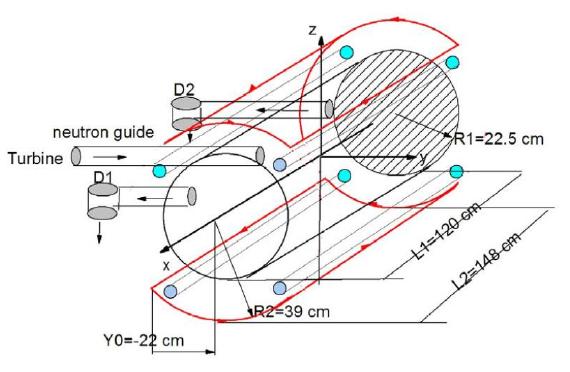}
\caption{\label{fig:trap} Scheme of the UCN chamber used in the experiment. 
Positions of entrance valve from the neutron guide and exit valves to detectors 
D1 and D2 are shown. Electric circuits inducing magnetic field in the trap are 
shown by red contours.  }
\end{center}
\end{figure}

The experiment  was carried out  
at the Research Reactor of the Institute Laue-Langevin (ILL), Grenoble, 
using the EDM beam-line of the UCN facility PF2.  
The vacuum chamber of PNPI spectrometer was used, the same that in previous experiments 
on  $n-n'$ transitions \cite{Serebrov1,Serebrov2}.  
The experimental set up  consisting of a neutron guiding system, 
the UCN storage trap with valves for filling and emptying, two UCN detectors and magnetic shielding 
is shown in Fig.~\ref{fig:trap}. The trap of $190\,\ell$ volume capable of storing about half million UCN 
has a form of cylinder with a length 120~cm and diameter 45~cm. Its inner surface is coated by beryllium.  
The trap is located inside a shield which screens the Earth magnetic field  (for more details, see Refs. \cite{Serebrov1,Serebrov2}). 

\begin{figure}[t]
 \begin{center}
\includegraphics[width=7cm]{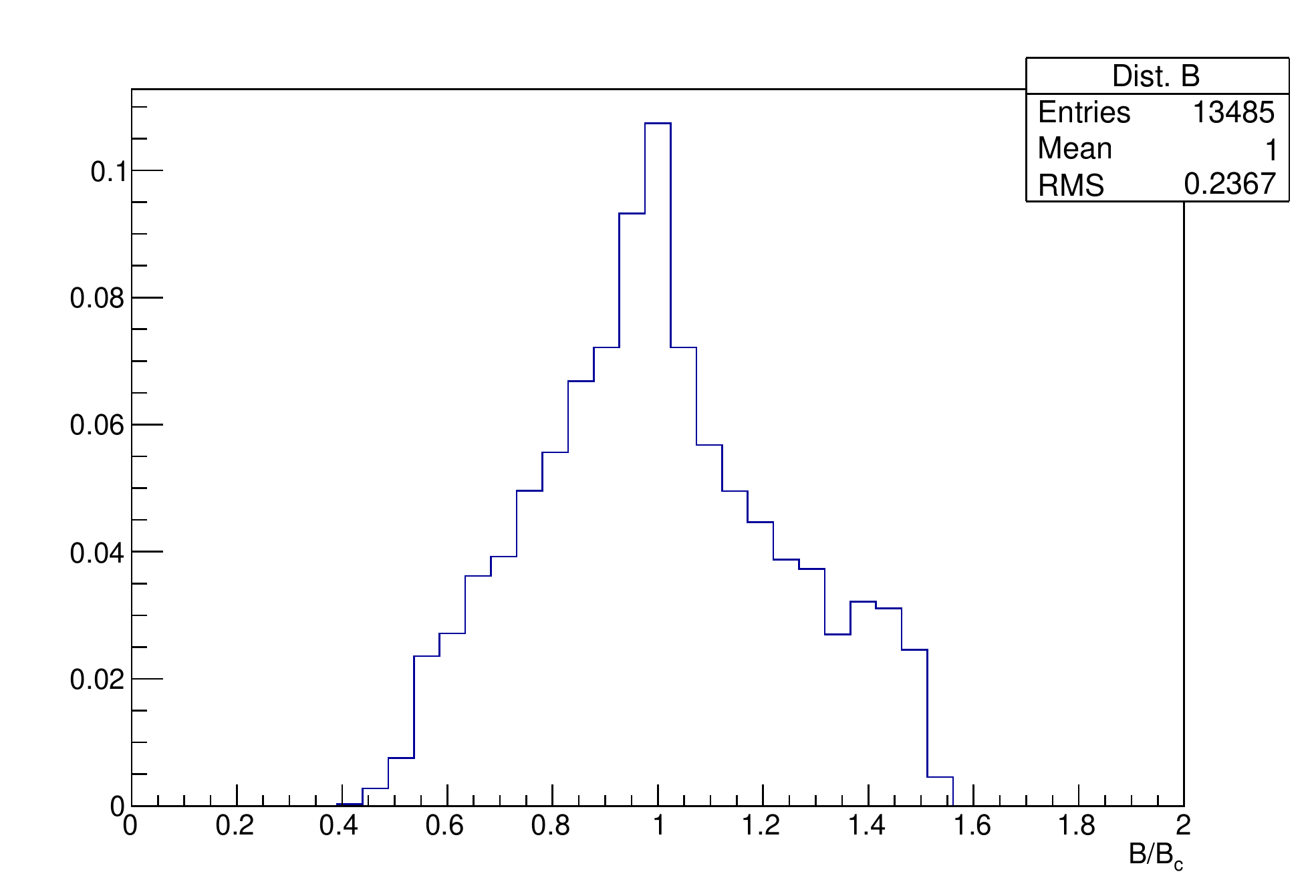}
\includegraphics[width=7cm]{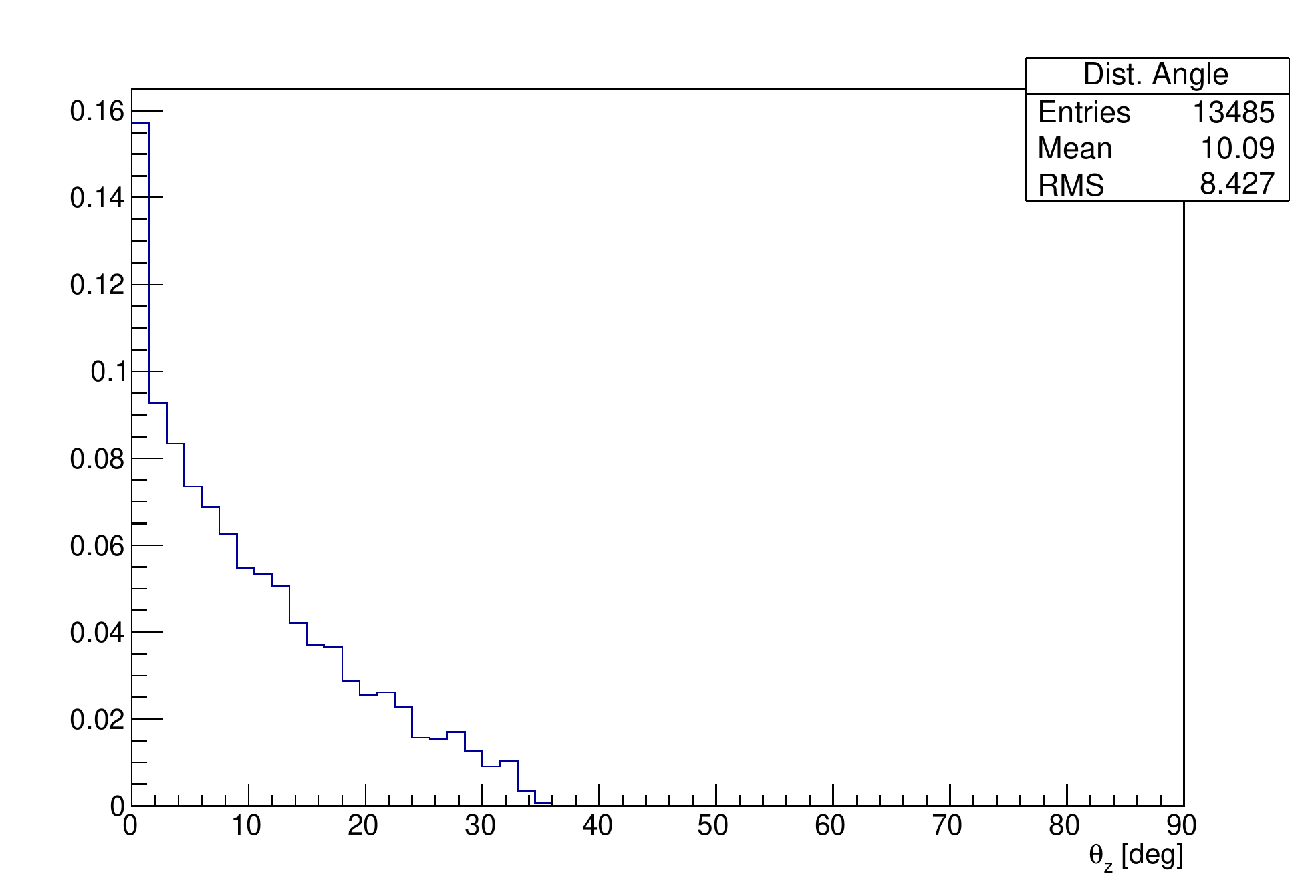}
\caption{\label{fig:dist} 
{\it Upper panel:} distribution of magnetic field around the central value $B_c$ inside the trap.   
{\it Lower panel:} distribution of the deviation angles  of vector $\vec{B}$ 
relative to vertical $z$ axis. }
\end{center}
\end{figure}

A controlled magnetic field $B$ was applied inside the trap using electric circuits placed on 
the top and bottom of the chamber (red contours in  Fig.~\ref{fig:trap}). 
For a given electric current in the circuits, the value of the induced magnetic field and 
its direction at each position inside the trap were calculated theoretically,  
by approximating the interior of the trap was by a cubic lattice, with coordinates $x,y,z$ from the 
center of the trap taken with 1~cm steps. 
The obtained results were also checked by with a magnetometer 
at center and some characteristic lateral parts of the chamber. 
The induced magnetic field $\vec{B}$ had practically vertical direction everywhere 
but its magnitude $B= \vert \vec{B} \vert$ was in-homogeneously distributed, 
varying from the value $B_c$ in the center by about $\pm 0.5\, B_c$ at peripheral regions. 
The distribution of magnetic field inside the trap is shown in Fig.~\ref{fig:dist}.  
 In the following,  for describing different experimental configurations we shall use 
 the central value of magnetic field $B_c$  induced by a proper electric current.  
Direction of magnetic field was periodically inverted by changing the direction of current.

The  scheme of the experiment is the following. 
Each measurement
consists of five phases: monitoring, filling, storage, emptying and background fixing. 
Typical time per measurement including the turbine waiting time is about 10 minutes.  
The monitoring phase is used to check the stability of the UCN flux from the reactor. 
After the entrance valve is open, neutrons flow into the trap via the UCN guide while the exit valves 
towards two detectors $D_1$ and $D_2$ remain open, 
and their counts during monitoring time $t_{\rm m} = 50$~s are used as estimators of the incident UCN flux.  
Then the exit valves are closed for a filling time of 100~s, 
after which the entrance valve is closed and the UCN are kept inside the trap 
for a holding time $t_{\rm s} = 250$~s.  Then the exit valves are reopen and the survived UCN are counted 
by two detectors during the emptying time of $150$~s. The background phase is for checking that 
no excess of neutrons remain inside the trap that could influence the subsequent measurement. 


In first three series of experiments ($B1$, $B2$, $B3$) the asymmetry $A_{\vec{B}}$ (\ref{AB}) 
was measured employing only large magnetic fields 
($B_c=0.21$~G,  $B_c=0.12$~G, and $B_c=0.09$~G respectively), 
repeating the cycles $\{ \vect{B} \} =
\{-\vect{B},+\vect{B},+\vect{B},-\vect{B};+\vect{B},-\vect{B},-\vect{B},+\vect{B}\}$, 
with the UCN holding time $t_{\rm s}=250$~s (signs $\pm$ correspond to the fields directed up and down). 
In last part of series $B4$  we used the measurement sequences 
$\{\vect{0}|\vect{B}\} =
\{\vect{0},+\vect{B},-\vec{B},\vec{0}; \vect{0},-\vec{B},+\vect{B},\vect{0} \}$, 
altering zero and non-zero values of magnetic field, again with $B_c=0.12$~G. 
For technical reasons, 
in these measurements only one detector ($D_1$) was used and the holding time 
was reduced to $t_{\rm s}=150$~s. Time gap between the series $B3$ and $B4$ was 
devoted to calibration measurements for testing possible systematic effects that
could render the detector counts sensitive to the magnetic field orientation, 
as e.g. influence of the alternating  current on the counting electronics. 
These measurements were performed with high statistics in continuous flow regime 
for $\{ \vect{B} \}$ mode at $B_c=0.09$~G and $B_c=0.12$~G,  
with entrance and both exit valves of the trap open during 200~s  
so that the neutrons entering the trap were finishing in one of two detectors  after a short diffusion time. 
Asymmetries $A_{\vec{B}}$ of neutron counts  $N_{\vec{B}}$ and $N_{-\vec{B}}$ acquired in this regime 
by both detectors $D1$ and $D2$  were compatible with zero, 
making it clear that the switching of magnetic fields had no influence. 
 After series $B4$,  calibration measurements were performed also in $\{\vect{0}|\vect{B}\}$ mode 
 with $B_c=0.16$~G and $B_c=0.21$~G, using only detector $D1$. 
 
 The averaged number of wall collisions $n_\ast$ and mean probabilities of $n-n'$ oscillation 
 (\ref{PD}) between collisions  were estimated via a Monte Carlo (MC) simulation. 
 It consisted of two steps which we briefly describe here. (The detailed description 
 will be given elsewhere \cite{Biondi}.)  
  
  First we estimated the average number of wall scatterings taking into account that 
 the initial velocity spectrum  of the UCN  \cite{UCN_spectra} entering the trap 
gradually degrades during the storage time due to regular neutron loses. 
 Therefore, a MC simulation was performed, first without considering the effects  
of $n-n'$ oscillation,  in order to obtain the 
mean value of free flight time $t_{\rm f} = \langle t \rangle$, its variance $\langle t^2 \rangle$, etc.   
In this way, the distribution of above values were computed 
by averaging them over the individual neutron trajectories in the trap, 
using well-known formulas \cite{Golub} for the UCN loses per scattering. 
The parameters were adjusted for reproducing the experimental data as characteristic 
time constants for the neutron counts during the trap filling,  UCN storage and the trap emptying. 
The obtained values are in good agreement with the parameters 
used in the previous experiments with the same trap \cite{Serebrov1, Serebrov2}. 
In this way, for a given storage time $t_{\rm s}$, 
we computed an average amount of wall scatterings $n_\ast$ 
that survived neutrons suffered starting from the moment they enter the trap in the  stage of filling, 
to the moment when they hit the detectors in the emptying phase.\footnote{Since $n-n'$ oscillation 
can take place not only during the UCN holding time $t_{\rm s}$ but also during filling and emptying of the trap, 
the effective exposure time can be estimated as $t_\ast = t_{\rm s} + 70$~s, 
as in previous experiments with the same trap \cite{Serebrov1,Serebrov2}. 
Let us remark that for us the relevant parameter is $n_\ast$ rather than $t_\ast$.  }
 Namely, we get $n_\ast = 2068\pm 18$ for  $t_{\rm s}=250$~s in $B1,B2,B3$ modes 
and $n_\ast = 1487 \pm 15$ for  $t_{\rm s}=150$~s in $B4$ mode  with one detector. 
The above error-bars related to fitting uncertainties 
introduce less than 1 \% systematic errors in the determination of 
$n-n'$ oscillation time. In the following, for deducing our limits in more conservative way, 
we use the lower values of $n_\ast$. 
   
\begin{figure}[t]
 \begin{center}
\includegraphics[width=8cm]{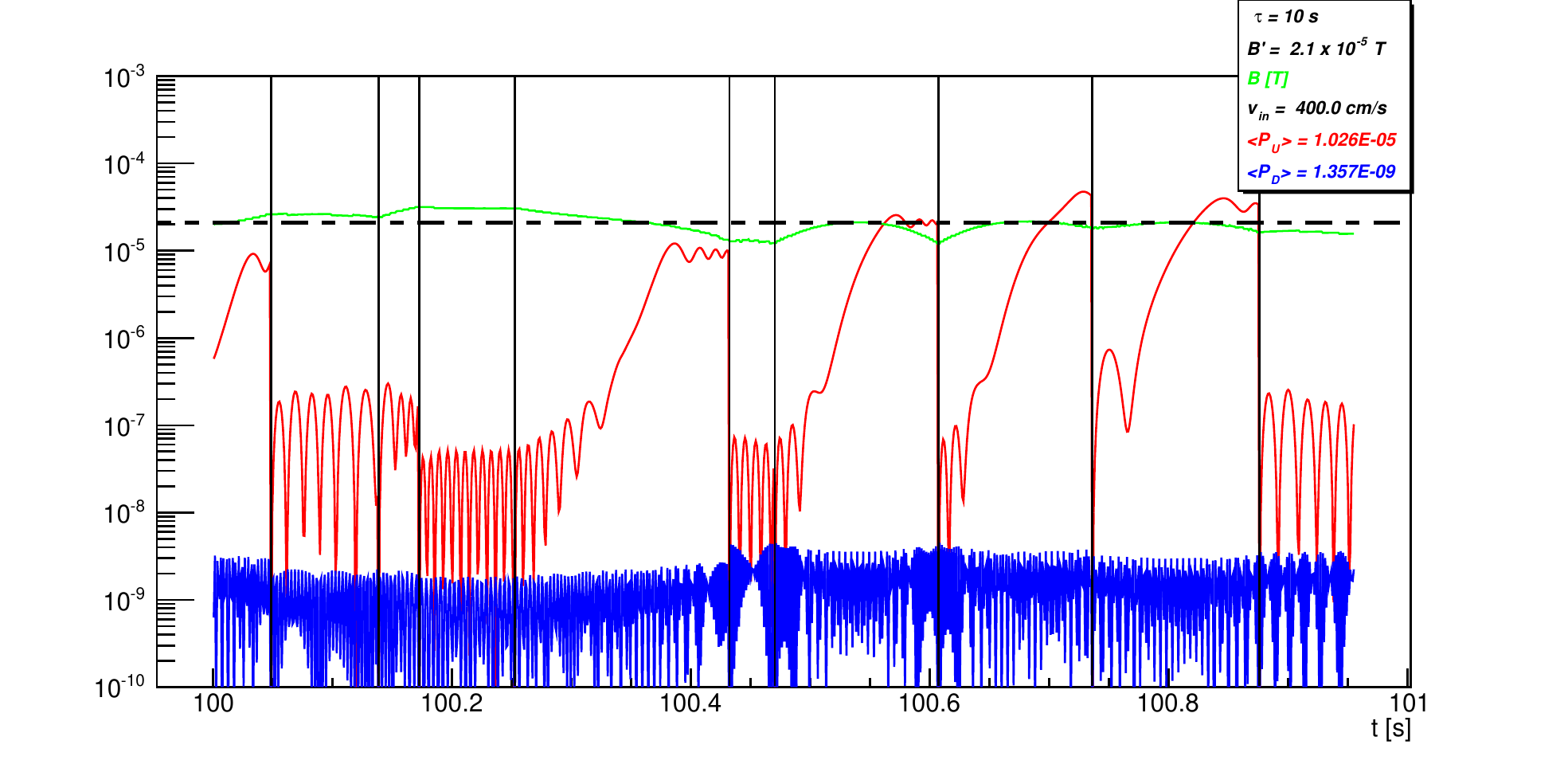}
 \includegraphics[width=8cm]{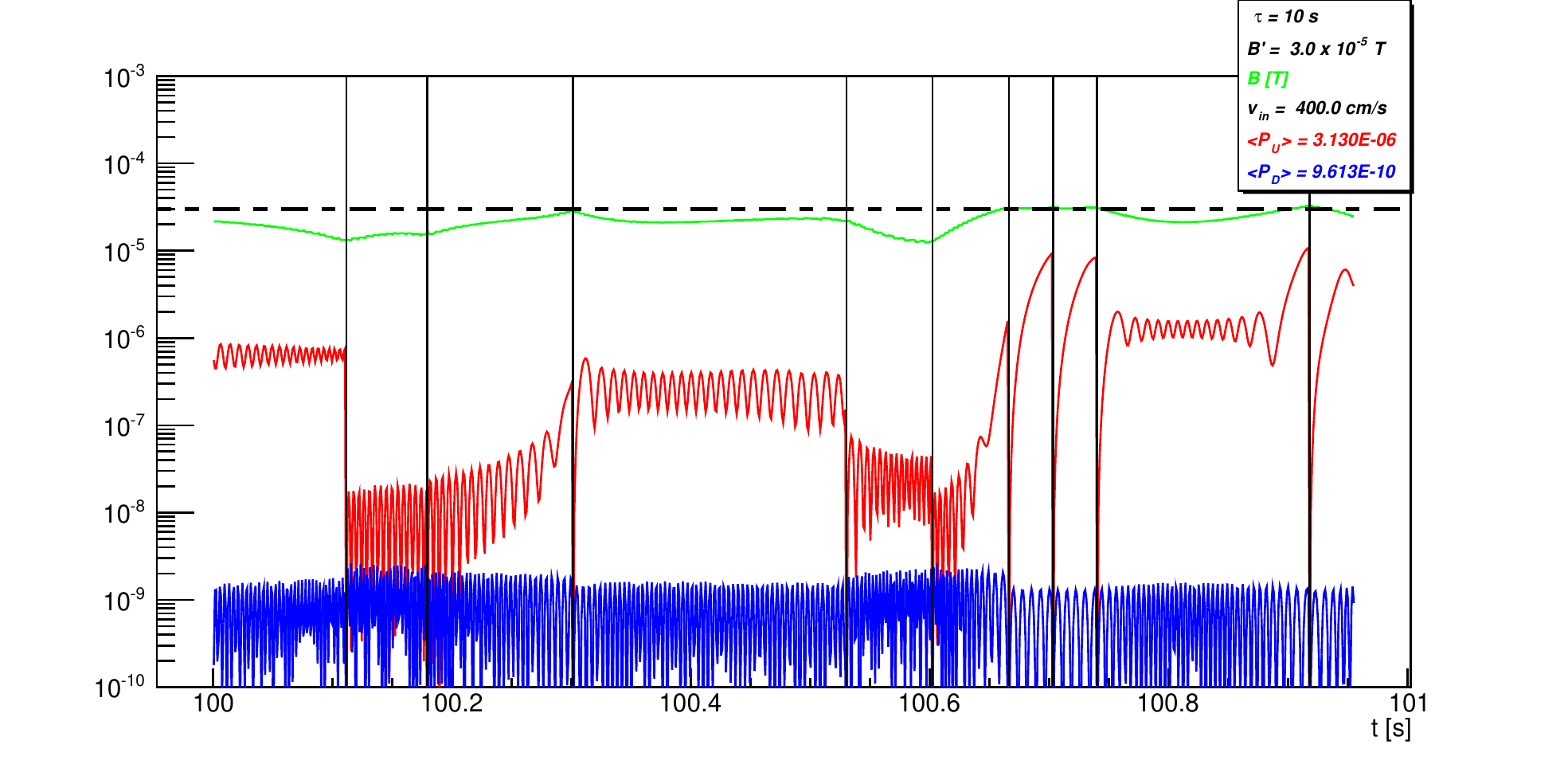}
\caption{\label{fig:simulation} 
Examples of evolution of $n-n'$ oscillation probabilities (\ref{maxmin}) between wall scatterings.  
Red and blue curves correspond to the cases $\beta=0$ and $\beta=\pi$. 
For definiteness, the central field is taken as $B_c=0.21$~G and $n-n'$ oscillation time is set as  $\tau=10$~s. 
At each wall scattering (marked by black vertical lines) 
the wave function resets to the pure neutron state.
For the sake of demonstration, mirror field (shown by horizontal dash lines) is taken as 
$B'=0.21$~G (upper panel) or $B'=0.30$~G (lower panel).  Green curves show the profile 
of applied magnetic field which the neutron crossed during its diffusion in the trap. 
}
\end{center}
\end{figure}

At second step, we computed  average oscillation probabilities (\ref{P}) over the neutron flight time. 
Given that the applied magnetic field in our experiment was not homogeneous, 
the empirical formula (\ref{formula}) cannot be used: for given central value $B_c$ in the trap, 
the neutrons during their diffusion can cross resonant values $B=B'$ with some probability 
if the mirror magnetic field has a value inside the distribution of magnetic field in the trap .  
Therefore, for each experimental series,  
we calculated the oscillation probability between wall scatterings as a function of $B'$ 
following the neutron trajectories in the trap. The interior of the trap was represented by 
a cubic lattice with 1~cm$^3$ elementary volumes, the magnetic field $\vec{B}$ was calculated 
in every node of this lattice, and in this way the distributions shown in Figs.~\ref{fig:dist} 
were obtained. In each elementary cm$^3$ volume the magnetic field was taken as constant, with 
a value obtained by averaging between 8 calculated values at its vertices.  
The Schr\"odinger equation (\ref{H-osc}) was numerically integrated along the neutron trajectories. 
Every neutron leaving an elementary cube with a given value of magnetic 
field with a wave vector $(\psi_n,\psi_{n'})$,  while crossing the adjacent cube was evolving 
 with the corresponding magnetic field. 
At each wall scattering the wave vector was reset to the pure state of neutron $(1,0)$.  
The typical evolution of $n-n'$ oscillation probabilities $P_{BB'}^{(\pm)}(t)$ 
(\ref{maxmin}) 
for a 1 second period of the neutron diffusion in the trap is shown 
in Fig. \ref{fig:simulation}. 

\begin{figure}[t]
 \begin{center}
 \includegraphics[width=8cm]{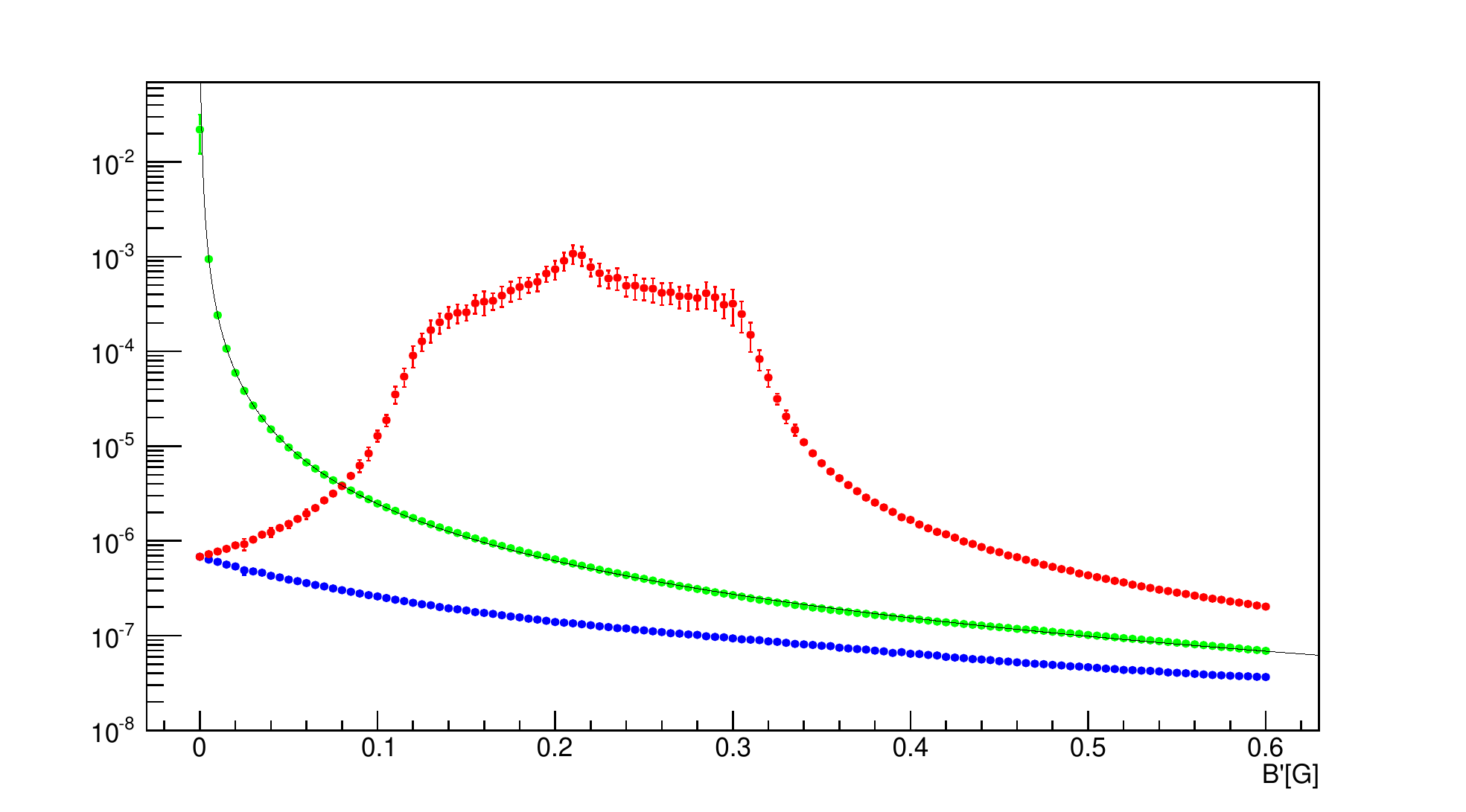}
 \includegraphics[width=8cm]{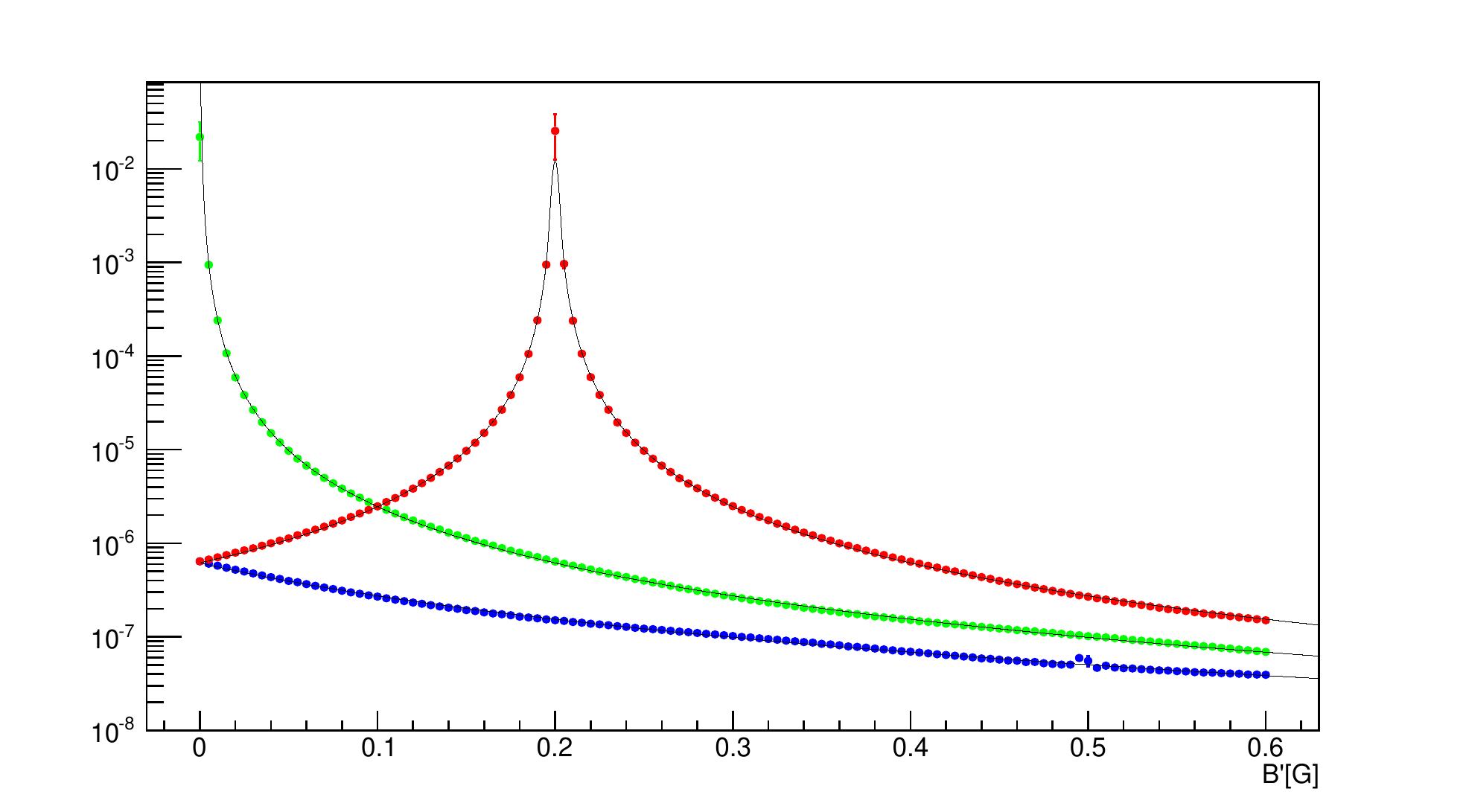}
\caption{\label{fig:MC} 
{\it Upper panel:} 
results of MC simulation for $S_+(B')$ (red points), $S_-(B')$ (blue points) and $S_0(B')$ (green points) 
for the case $B_c = 0.21$~G. 
{\it Lower panel:}  MC results in the case of homogeneous magnetic field $B=B_c$. 
Black solid curves show the results obtained via empirical formula (\ref{formula}). } 
\end{center}
\end{figure}

In this way, via MC simulations we computed mean values 
$\ov{P}_{BB'}^{(\pm)}$ of the probabilities (\ref{maxmin})  between wall scatterings, 
averaged over distribution of the neutron flight time $t$ 
and distribution of the magnetic field $B$ in the trap for a given value of $B_c$.  
In addition, we computed also the mean oscillation probability $\ov{P}_{0B'}$ 
for the case when no magnetic field was applied, $B=0$. 
%
\beqn{SBc}
&& \ov{P}_{BB'}^{(\pm)}  =
\left\langle \frac{\sin^2[(\om \mp \om')t]}{\tau^2 (\om \mp \om')^2} \right\rangle_{t,B} = 
\left(\frac{1\, {\rm s}}{\tau}\right)^2 \! S_{\pm}(B')  \, ,  \nonumber \\
&& \ov{P}_{0B'} =  \left\langle \frac{\sin^2(\om')t }{\tau^2 \om^{\prime2}} \right\rangle_t
= \left(\frac{1\, {\rm s}}{\tau}\right)^2 \! S_{0}(B') \, . 
\eeqn 
In upper panel of Fig. \ref{fig:MC} we show $S_{\pm}(B')$ and $S_0(B')$  
as functions of mirror field strength $B'$. As we see, these factors correspond to 
mean values of the respective probabilities normalized to $n-n'$ oscillation time $\tau=1$~s. 

For checking the consistency of our MC simulation, we also computed average oscillation 
probabilities (\ref{SBc}) 
in the case of homogeneous magnetic field in the same trap (lower panel of Fig. \ref{fig:MC}). 
As we see, the results of MC simulation 
perfectly coincide with the results (black solid curves) 
obtained via the empirical formula (\ref{formula}) with the corresponding MC values 
of $t_{\rm f}  = \langle t \rangle$ and $\sigma^2_{\rm f} = \langle t^2 \rangle - t_{\rm f}^2$.  
As we see from Fig. \ref{fig:MC}, inhomogeneous profile of magnetic fields has 
certain advantages:  the function $S_-{B'}$ in homogeneous magnetic field has maximal sensitivity 
at the resonance, $B\approx B'$ than in inhomogeneous case, but in the latter case 
it covers much wider range of $B'$.

\section{Data analysis}

Different datasets $B1$, $B2$, $B3$ and $B4$ were independently analyzed 
and the values of asymmetries $A_{\vec{B}}$ (\ref{AB}) were computed  via comparing 
the neutron counts $N_{\vec{B}}$ and $N_{-\vec{B}}$ between subsequent measurements. 
For each individual measurement under given applied field $\vec{B}$ 
we take $N_{\vec{B}}$ as a sum of counts of both detectors, 
$N_{\vec{B}}=N^{(1)}_{\vec{B}} +N^{(2)}_{\vec{B}}$, while the count ratios between two detectors
$N^{(1)}_{\vec{B}}/N^{(2)}_{\vec{B}}$  were also controlled as the stability check. 
Assuming Poisson statistics, the errors can be estimated as $\Delta N_{\vec{B}} = \sqrt{N_{\vec{B}}}$. 
For eliminating the effects of drift, in our analysis we use the values of $A_{\vec{B}}$ averaged within 
the measurement octets $\{\vect{B}\}$ for series $B1,B2,B3$ and $\{\vect{0}|\vect{B}\}$ for series $B4$.  
Hence, a cycle $\{\vect{B}\}$ of 8 measurements yield an average of 4 measured values for 
$A_{\vec{B}}$  while 8 measurements  of cycle  $\{\vect{0}|\vect{B}\}$ yield 2 values of $A_{\vec{B}}$ 
and 2 values of $E_B$. In this way, for series  $B4$ also the values $E_B$ (\ref{EB}) were computed.  
In addition, the UCN counts $M_{\vec{B}}$ and $M_{-\vec{B}}$
in the monitoring phase were also controlled, 
and detector-to-monitor normalized asymmetries $A_B^{\rm nor}$  between the 
ratios $(N/M)_{\vec{B}}$ and $(N/M)_{-\vec{B}}$, and analogously  $E_B^{\rm nor}$. 
In this way,  the average values of $A_{\vec{B}}$ and $E_B$ 
obtained in each measurement series, 
were transformed into mean probabilities $\ov{\cD}_{\vec{B}}=A_{\vec{B}}/n_\ast$ (\ref{DB})
 and $\ov{\Delta}_B = E_B/n_\ast$ (\ref{DeltaB}) 
taking the mean amount of wall scatterings computed via MC simulations as 
$n_\ast = 2050$ for configurations $B1,B2,B3$ and $n_\ast=1472$ for $B4$. 

\begin{figure}[t]
 \begin{center}
\includegraphics[width=9cm]{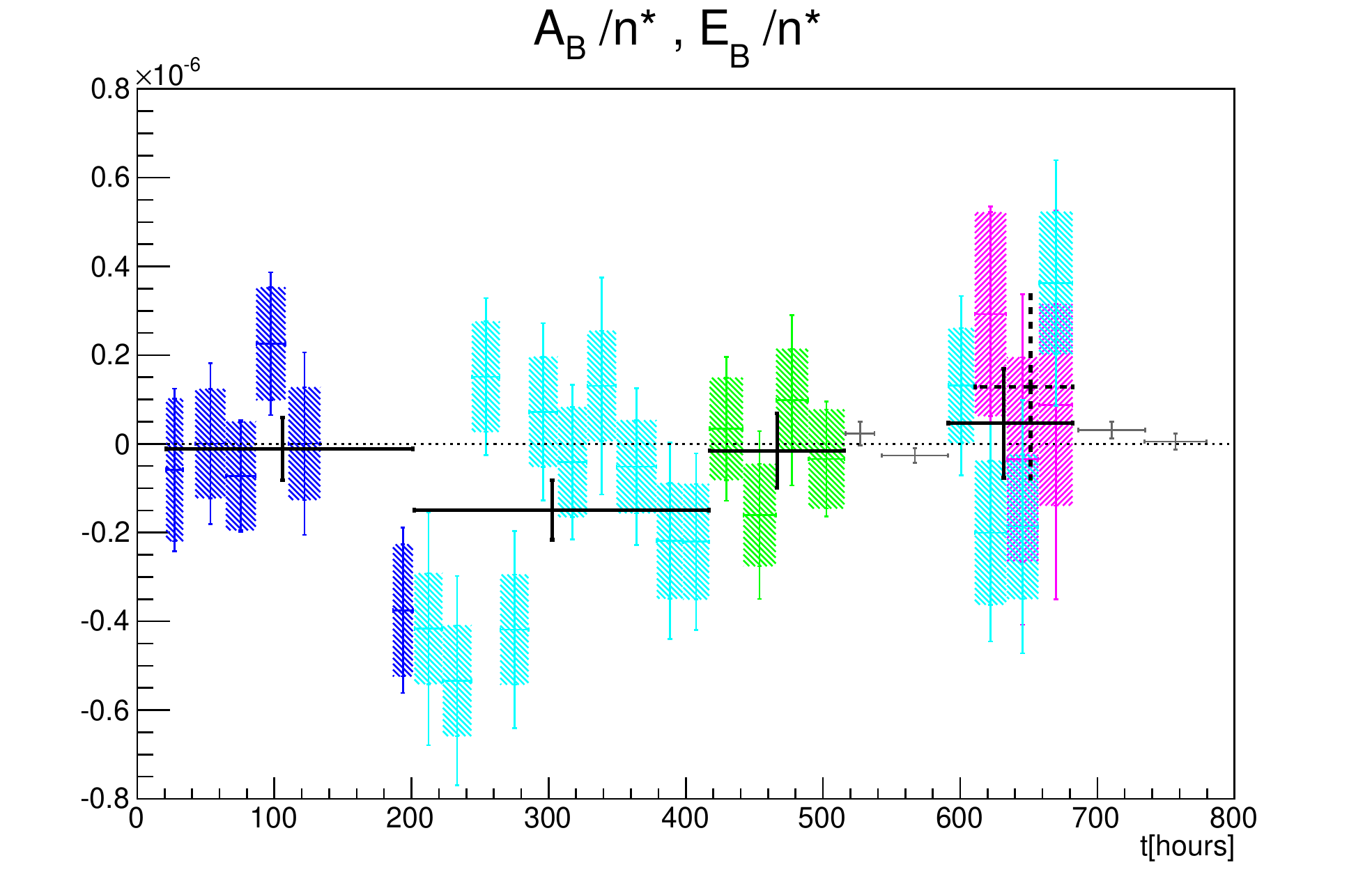}
\includegraphics[width=9cm]{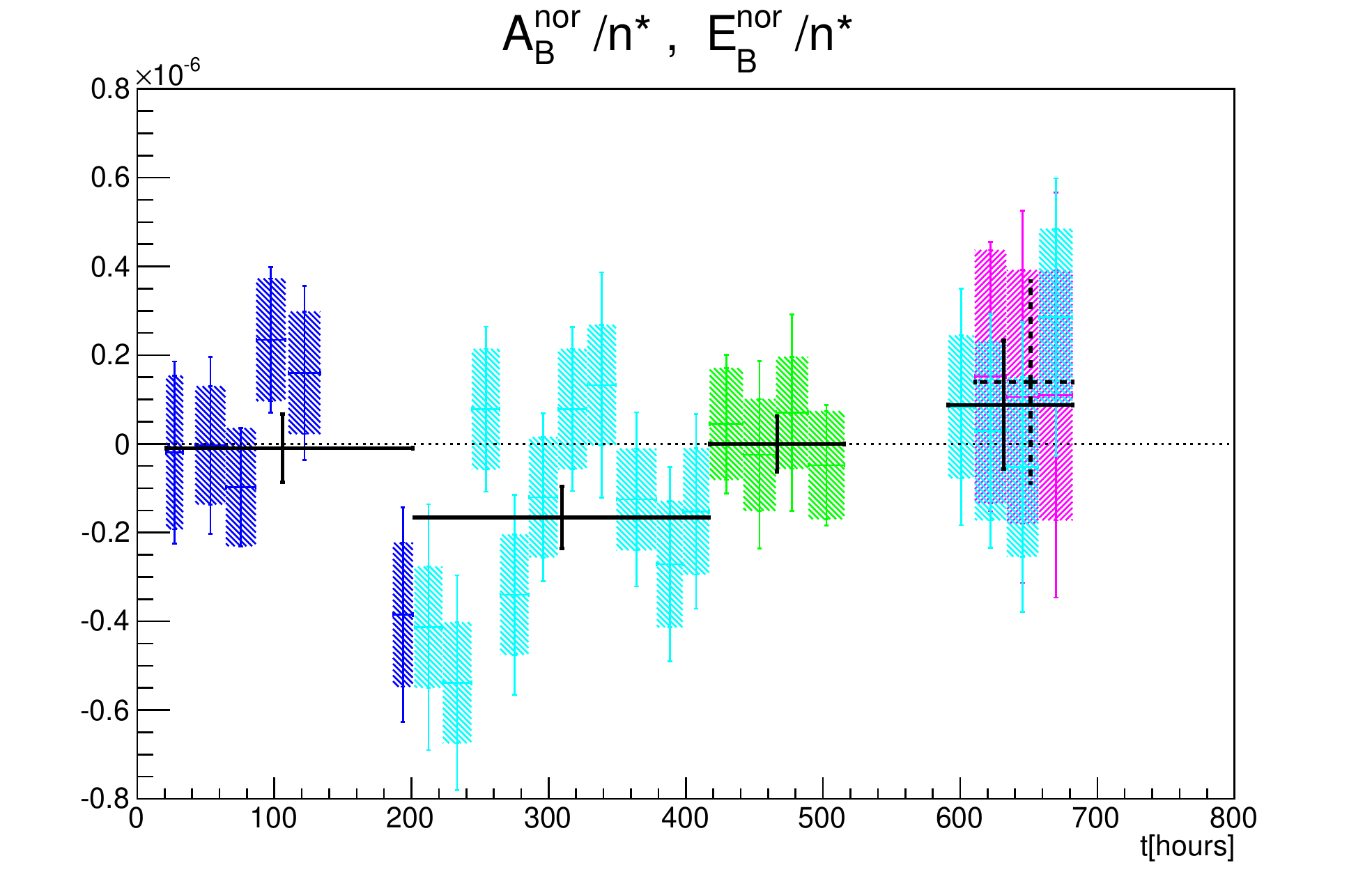}
\caption{\label{fig:custom} 
Binned results for $\ov{\cD}_{\vec{B}}$ {\it (upper panel)} and $\ov{\cD}_{\vec{B}}^{\rm nor}$ {\it (lower panel)} 
from the data acquired in configurations $B1$ (blue), $B2$ (cyan), $B3$ (green) and $B4$ (again cyan). 
Shaded  squares correspond to expected statistical errors, while the data dispersion 
in bins are indicated by longer error bars. 
Magenta squares and error bars indicate the results for $\ov{\Delta}_B$ and  $\ov{\Delta}_B^{\rm nor}$ from 
series $B4$. Grey crosses in upper panel correspond to calibration series.}
\end{center}
\end{figure}

The results obtained per each measurement series 
are shown in Table~\ref{tab:1} and also on Fig.~\ref{fig:custom} where 
the measured asymmetries are combined in bins of comparable size. 
The first column of Table \ref{tab:1} indicates  the values measured in a given series, 
and corresponding amount $N_{\rm oct}$ 
of the measurement octets  $\{\vect{B}\}$ or $\{\vect{0}|\vect{B}\}$.   
(Let us remind that  for canceling the effects of drift, each data unit 
is taken as value of $A_{\vec{B}}$ or $E_B$ averaged within a given octet of measurements, 
$\{\vect{B}\}$ or $\{\vect{0}|\vect{B}\}$.   
Thus for a constant fit  the amount of degrees of freedom per each series is $N_{\rm oct}-1$. 
The second column of Table~\ref{tab:1} shows  $\ov{\cD}_{\vec{B}}$ and $\ov{\Delta}_B$ 
deduced respectively from the average values  of $A_{\vec{B}}$ and $E_B$ in each series,  
and the expected statistical errors (with statistical fluctuation for every count $N$ taken as $\sqrt{N}$).   
However, the corresponding values of $\chi^2/{\rm d.o.f.}$ (in parenthesis)  are too large 
which indicates that these fits are not that good. 
Third column shows the mean values of $A_{\vec{B}}$ and $E_B$ and
 respective variances obtained directly the distribution of their measured values in each series.    
As we see, the central values in third column are consistent to that of second column, however 
the error bars are larger. In fact, the latter errors well coincide with the respective  
statistical errors  enlarged by the respective value of $\sqrt{\chi^2/{\rm d.o.f.}}$. 

 \begin{table}[t]
\begin{center}
\begin{tabular}{cccc}
\colrule
               & Stat. $ [\times 10^{-8}]$    & Dist. $ [\times 10^{-8}]$ \\
\colrule
$A_{B1}/n_\ast$ [74]          & $-1.59 \pm 5.40$ (1.57)  & $-1.12 \pm 7.09$ \\
 $A_{B1}^{\rm nor}/n_\ast$ [74] & ~ $0.43 \pm 5.89$ (1.73)   & ~ $0.99 \pm 7.67$ \\
 \colrule
$A_{B2}/n_\ast$ [124]       & $-14.8 \pm 3.90$ (2.90)  & $-14.9 \pm 6.60$  \\
 $A_{B2}^{\rm nor}/n_\ast$ [124]    & $-16.5 \pm 4.24$ (2.84)  & $-16.6 \pm 6.90$   \\
\colrule
$A_{B3/n_\ast}$ [57]       & $-0.03 \pm 5.79$ (1.92)  & $-1.54 \pm 8.39$   \\
 $A_{B3}^{\rm nor}/n_\ast$ [57]   & ~ $1.93 \pm 6.32$ (1.83)    & ~ $0.96 \pm 9.07$   \\
\colrule
$A_{B4/n_\ast}$ [43]          &$4.18 \pm 7.47$ (2.20) & $4.57 \pm 12.1$  \\
 $A_{B4}^{\rm nor}/n_\ast$ [43]     &$8.61 \pm 9.28$ (2.50)  & $8.67 \pm 14.3$  \\
$E_{B4}/n_\ast$ [28]       &$13.0 \pm 13.0 $ (2.20) & $12.8 \pm 20.4$   \\
 $E_{B4}^{\rm nor}/n_\ast$ [28]  &$13.7 \pm 13.7 $ (1.94) & $13.7 \pm 22.4$  \\
\colrule
\colrule
\end{tabular}
\caption{ Results for $\ov{\cD}_{\vec{B}}$ and 
$\ov{\Delta}_B$ obtained 
from  the average values of $A_B$ and $E_B$ measured in respective experimental cycles 
taking into account only statistical errors (in parenthesis the quality of constant fit  
($\chi^2/{\rm d.o.f.})$ is shown). 
Last column shows the mean values and variance reconstructed directly from the distribution 
of experimental data. }
\label{tab:1}
\end{center}
\end{table}

Wide black crosses  in~Fig. \ref{fig:custom}  
 show the mean values of $A_{\vec{B}}/n_\ast$ and $A_{\vec{B}}^{\rm nor}/n_\ast$ 
and respective errors obtained per each series obtained directly from the distribution of the measured values,  
in correspondence to third column of Table~\ref{tab:1}.  
The dashed black  crosses show the same for $E_B/n_\ast$ and $E_B^{\rm nor}/n_\ast$,    
and grey crosses show results of calibration measurements. Shaded squares show mean values 
per each bin and statistical errors, while the larger error-bars indicate the 
data dispersion in each bin. 

In ideal situation, our measurements  should have no systematical uncertainties  
since the regular neutron loses should not affect the values of $A_{\vec{B}}$ or $E_B$ 
measured in the same experimental conditions. 
Hence,  in the absence of $n-n'$ oscillations one expects that 
the values $A_{\vec{B}}$ and $E_B$ should be consistent with zero within statistical errors. 
Table \ref{tab:1} shows that average values measured in each experimental series 
are consistent with null hypothesis  within $1\sigma$ statistical errors except that of largest 
series $B2$,  comprising over 200 hours of continuous measurements, 
where the values $A_{\vec{B}}$ and $A_{\vec{B}}^{\rm nor}$ both show about $4\sigma$ 
deviation from zero. 
On the other hand, the quality of constant fits  presented in Table \ref{tab:1} is not that good 
(namely, $\chi^2/{\rm d.o.f.}=2.9$ for series $B2$)   
 which means that some unaccounted external factors were influencing our measurements.   


As one can see on  Fig.~\ref{fig:custom},  
the results of series $B2$ (and perhaps also of series $B1$) 
show a strong dispersion between different bins which should be a main reason for bad constant fit. 
(In contrast, the results of series $B3$ show no significant dispersion between different bins.) 
On the other hand, even with enlarged error bars (third column of Table~\ref{tab:1}) both 
values $A_{\vec{B}}$ and $A_{\vec{B}}^{\rm nor}$ of series $B2$ still have about $2.3\sigma$ deviation 
from zero. The same result can be obtained by averaging between the bins of series $B2$ 
with enlarged error bars which shows that this discrepancy is pretty robust against the methods 
of the analysis.   
In principle, this situation could be interpreted as a signal of $n-n'$ oscillation in the presence of 
mirror magnetic field $B'\sim 0.1\div0.2$~G with a direction varying in time. 
However, duration and acquired statistics of our experiment is not enough to draw such a far going conclusion. 
Therefore, we take more conservative attitude, and in the following we use the mean values and
 their variances obtained directly the distribution of the measured values of $A_{\vec{B}}$ and $E_B$ 
 per each series,  shown in third column of Table~\ref{tab:1}.  
In addition, we average between the results of series $B2$ and $B4$ performed under the same 
magnetic field $B_c=0.12$~G and thus obtain 
\beqn{average04} 
&& \ov{\cD}_{\vec{B}\vec{B}'}[B_c=0.12~{\rm G}] = (-10.4 \pm 5.80) \times 10^{-8}, \nonumber \\
&&  \ov{\cD}_{\vec{B}\vec{B}'}^{\rm nor}[B_c=0.12~{\rm G}] = (-11.8 \pm 6.20) \times 10^{-8} \, . 
 \eeqn
This averages have less than $2\sigma$ deviation from zero and thus can be used for 
setting 95 \% C.L. on the oscillation time $\tau_\beta$ as a function of mirror magnetic 
field $B'$ assuming that the direction of the latter is not time variable. 




\section{Results for $n-n'$ oscillation parameters}

\begin{figure}[t]
 \begin{center}
\includegraphics[width=8cm]{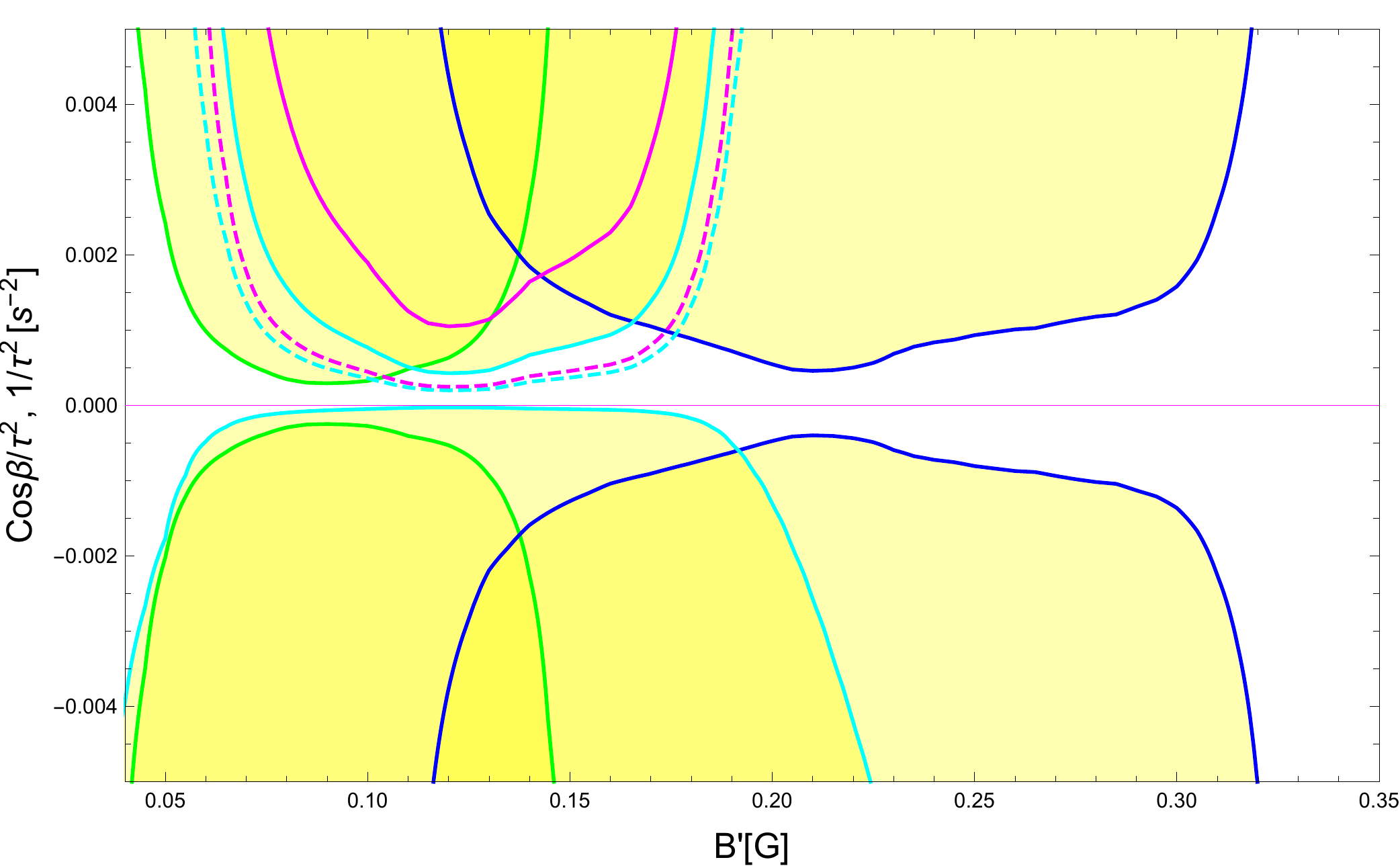}
\caption{\label{fig:limits} Exclusion regions for  $1/\tau^2$ and $\cos\beta/\tau^2$ 
extracted from our measurements of $E_B$  and $A_B$. Curves of different 
colors, corresponding to the colors of bins in Fig. \ref{fig:custom}
confine regions excluded by measurements at different values of $B_c$.} 
\end{center}
\end{figure}

Experimental values of $E_B/n_\ast = \ov{\Delta}_{BB'}$ and 
$ A_{\vec{B}}/n_\ast = \ov{\cD}_{\vec{B}\vec{B}'} = \ov{D}_{BB'}\cos\beta$ 
shown in Table~\ref{tab:1} can be transformed into the $n-n'$ oscillation 
parameters $\tau^2$ and $\tau^2/\cos\beta $ via Eqs. (\ref{SBc}):   
\beqn{tau-taubeta} 
&& \frac{1}{\tau^2} \big[{\rm s}^{-2}\big]
= \ov{\Delta}_{BB'}^{\rm exp} \left[\frac{S_+(B') + S_-(B')}{2}  - S_0(B') \right]^{-1} \nonumber \\
&& \frac{\cos\beta}{\tau^2} \big[{\rm s}^{-2}\big]
= \ov{\cD}_{BB'}^{\rm exp} \left[S_+(B') - S_-(B') \right]^{-1}
\eeqn 
The obtained results are shown in Fig.~\ref{fig:limits}. 
Dash magenta curve shows values of $1/\tau^2$ as function of $B'$ reproduced from 
central values of $E_B/n_\ast$ in Table \ref{tab:1}, while 
solid magenta curve corresponds to 95 \% C.L upper limit on $1/\tau^2$ 
obtained via taking into account respective error-bars of third column.  
Dash cyan curve shows central values of $\cos\beta/\tau^2$ obtained from 
central value of $\ov{\cD}_{BB'}$  in (\ref{average04}), an average result between the 
measurements  $B2$ and $B4$ with about $2\sigma$ deviation from zero, 
while solid cyan contours confine corresponding 95 \% C.L. area.  
(let us remind that $\cos\beta$ can be positive or negative; 
here $\beta= 0$ corresponds to mirror magnetic field directed to the Earth center.) 
Blue and green solid contours show 95 \% C.L. limits on  $\cos\beta/\tau^2$ deduced 
from results for $A_B/n_\ast$ for series $B1$ and $B3$, third column of Table \ref{tab:1}. 

\begin{figure}[t]
\includegraphics[width=8.5cm]{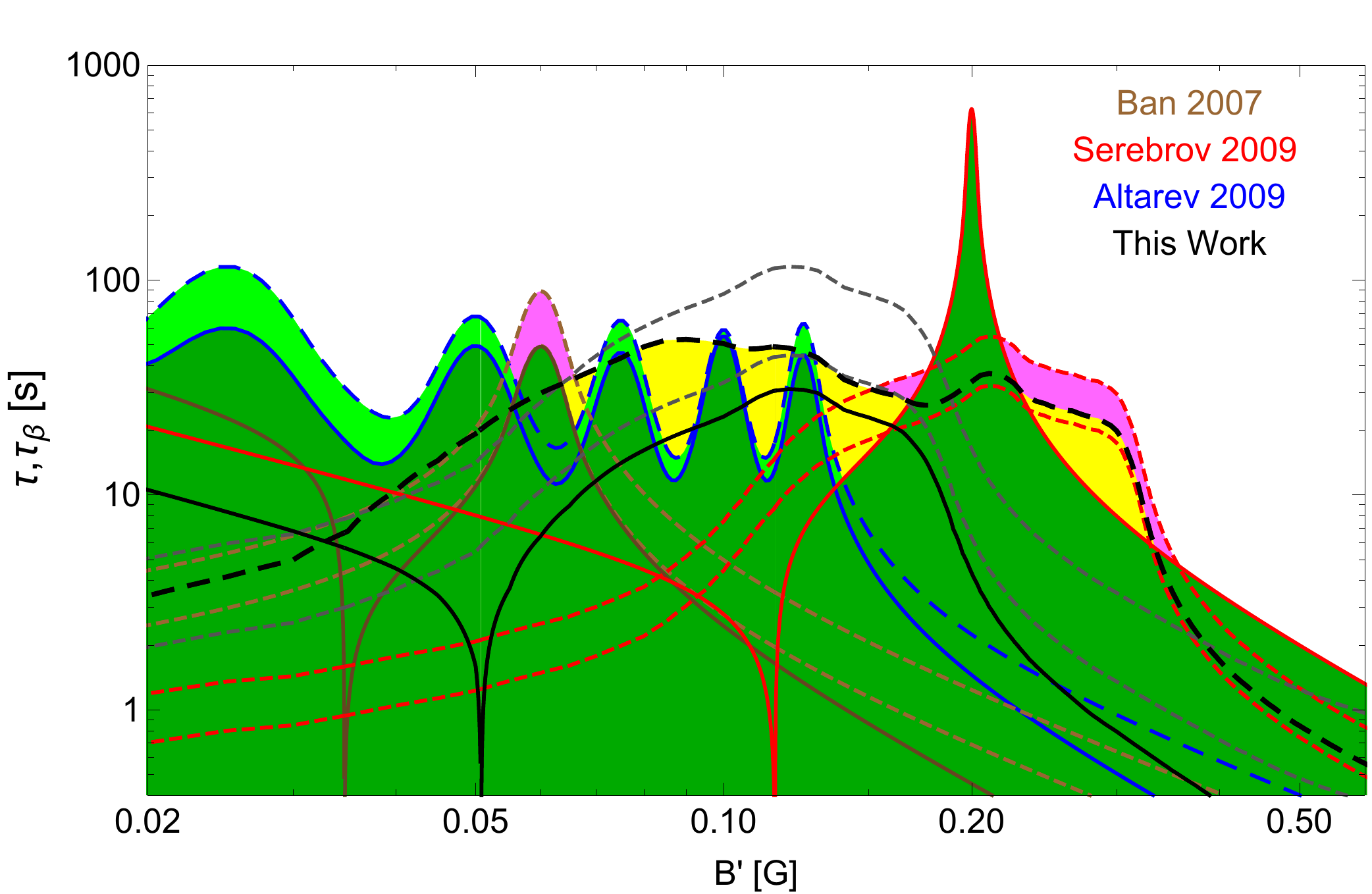}
\includegraphics[width=8.5cm]{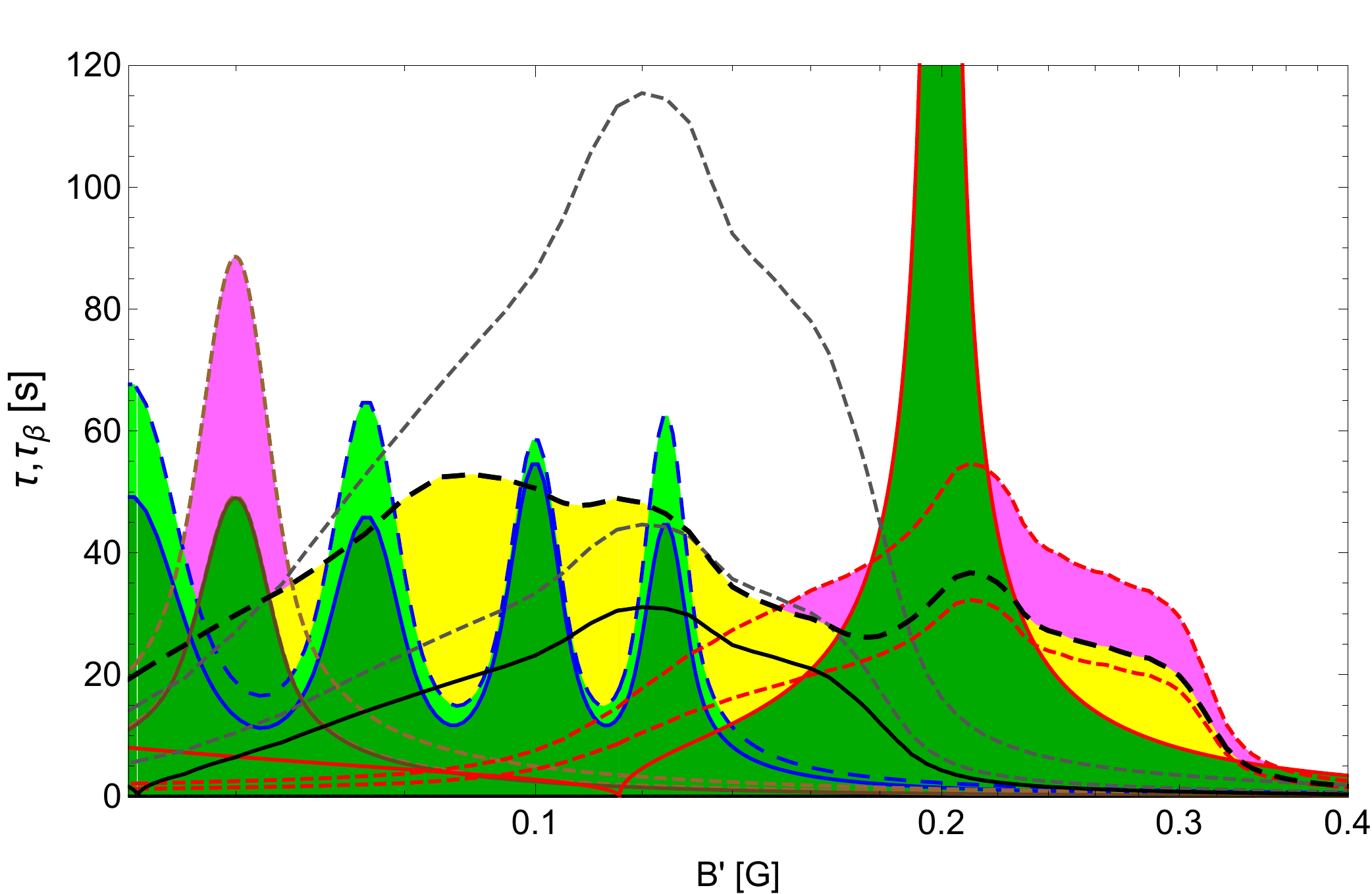}
\caption{\label{fig:global}
{\it upper panel:} our 95 \% C.L. lower limits  on $\tau$ (black solid) and $\tau_\beta$ (black dashed).  
 The parameter areas excluded by the previous experiments are shaded in dark green for $\tau$ and 
 light green for $\tau_\beta$ while the yellow shaded area 
corresponds to new exclusion regions of this work.
 Solid red curve corresponds to 99 \% C.L. lower limit on $\tau$ 
 from experiment \cite{Serebrov2}. 
 while dotted red contours confine $2\sigma$ region of $\tau_\beta$ 
 for $5\sigma$ $A_{\vec{B}}$ anomaly in vertical magnetic fields with $B_c=0.21$~G \cite{Serebrov2,Nesti}. 
 Wavy blue curves show 95 \% C.L. lower limits on $\tau$ (solid) and $\tau_\beta$ (dashed) from Ref. \cite{Altarev}. 
 Solid magenta curve corresponds to 95 \% C.L. lower limit on $\tau$ from experiment \cite{Ban} while 
 the dotted magenta curves confine $2\sigma$ region of $\tau_\beta$ for $3\sigma$ $A_{\vec{B}}$ anomaly 
 in the same experiment. 
 The parameter areas relevant for these anomalous deviations that are not excluded 
 are shaded in pink. 
 {\it Lower panel:} the same in linear scale, as the blow up of the exclusion regions 
 covered by our experiment. 
 }
\end{figure}

In Fig. \ref{fig:global} we show results of global fit of our experimental data, 
95 \% C.L. lower limits  on $\tau$ (black solid) and $\tau_\beta$ (black dashed), 
and confront them with the results of previous experiments.  

In particular, the first experiment \cite{Ban} searching for $n-n'$ oscillation 
compared the UCN losses between measurements in zero magnetic field $B=0$ 
and non-zero field $B=0.6$~G while the direction of the latter was altered 
between vertical up (+$\vec{B}$) and down  (-$\vec{B}$). 
No significant deviation from zero was found in the value of $E_B$ (\ref{EB}). 
The corresponding 95 \% C.L. lower limit on $\tau$ as a function of $B'$ 
is shown by solid magenta curve in Fig.~\ref{fig:global}. 
On the other hand,  the data reported in Table~1 of Ref. \cite{Ban},   
indicate towards a non-zero asymmetry 
between the counts $N_{\vec{B}}$ and $N_{-\vec{B}}$ in all measurement series 
with different storage times (see also Table~1 of Ref. \cite{More}).  
In overall, this corresponds to $3\sigma$ deviation of $A_{\vec{B}}$ (\ref{AB}) 
from zero which can be interpreted via $n-n'$ oscillation in the presence of mirror magnetic field 
Ref. \cite{More}. The  $2\sigma$ allowed area corresponding to this deviation 
is shown by dash magenta contours. 

The experiment  \cite{Altarev} was preformed to search $n-n'$ oscillation 
in the presence of mirror magnetic field, by varying the values of applied magnetic field (vertical) 
from 0 to 0.125~G with a step of 0.025~G and also altering its direction from up to down.  
In  Fig.~\ref{fig:global} we show 95 \% C.L. lower limits on $\tau$ (blue solid)  
and $\tau_\beta$ (blue dashed) as functions of $B'$ obtained via fitting the data 
reported in Fig.~1 or Ref. \cite{Altarev}.  Overally, these limits exclude $\tau < 12$~s 
and $\tau_\beta < 15$~s for any $B'$ at 95 \% C.L. for any $B'$ in the interval from 
$(0.02 \div 0.13)$~G. 
For smaller mirror fields, $B' < 0.02$~G, the lower limits on $\tau$ or $\tau_\beta$ are stronger, 
approaching 100~s or so, as obtained in Ref. \cite{Nesti} 
by combining  the results of  Ref. \cite{Altarev}  with that 
of Ref. \cite{Serebrov1} performed under the magnetic field $B=0.02$~G.  

The experiment \cite{Serebrov2} used the larger 
magnetic field, $B=0.2$~G, horizontally directed, and zero field, $B=0$. 
99 \% C.L. lower limit on $\tau$ as a function of $B'$ resulting from these measurements 
is shown by red solid curve in Fig. \ref{fig:global}. 
However, in the same experiment \cite{Serebrov2} 
the measurements  performed with vertical magnetic field $B\simeq 0.2$~G, 
has shown substantial asymmetry between the counts $N_{\vec{B}}$ and $N_{-\vec{B}}$. 
The detailed analysis of these experimental data performed in Ref. \cite{Nesti} for the asymmetry 
$A_{\vec{B}}$ indicate towards $5.2\sigma$ deviation from the null hypothesis, 
which can be interpreted as a signal of $n-n'$ oscillation in the background of mirror magnetic field  
$B' \simeq 0.1\div 0.2$~G.  
Let us remark that in Ref. \cite{Nesti} the consequences of this anomaly 
for $\tau_\beta$ as a function of $B'$ was deduced assuming the homogeneity of the applied 
magnetic field and using the profile of functions $S_\pm(B')$ (\ref{SBc}) 
calculated via the analytic formula (\ref{formula}), as shown on lower panel of Fig. \ref{fig:MC}. 
However, as we realized while performing our experiment in the same conditions, 
the magnetic field distribution was rather wide.  
Therefore, in this paper we recalculated $\tau_\beta$ as a function of $B'$ using properly 
the functions $S_\pm(B')$ computed via our MC simulations (upper panel of Fig.~\ref{fig:MC}). 
The red dashed contours n Fig.~\ref{fig:global} confine the obtained $2\sigma$ area  corresponding 
to this anomaly.

In Fig. \ref{fig:global} parameter areas excluded by the previous experiments in overall 
are shaded in dark green for $\tau$ and  light green for $\tau_\beta= \tau/\sqrt{\cos\beta}$ 
while the yellow shaded areas correspond to new regions excluded in this work. 
The pink shaded areas correspond to parameter regions relevant for the  anomalies  
in $A_{\vec{B}}$ from Ref. \cite{Ban} (see also Ref. \cite{More}) and 
Refs. \cite{Serebrov2,Nesti} which still remain allowed 
by the present experimental limits. 


\section{Conclusions and outlook}

The aim of this experiment was to test the results of experiment \cite{Serebrov2} 
with vertical magnetic fields at $B_c=0.21$~G, 
showing $5\sigma$ deviation from the null hypothesis \cite{Nesti} that could be 
interpreted as a signal for $n-n'$ oscillation in the presence mirror magnetic field 
$B'\sim 0.1$~G, and in fact  imply an {\it upper} limit on $n-n'$ oscillation time 
$\tau \leq \tau_\beta < 57$~s at 95 \% C.L. 
 (the corresponding $2\sigma$ region for $\tau_\beta$ as a function of $B'$ 
is confined between dotted red contours in Fig.~\ref{fig:global}). 
The results of other experiments restrict the parameter space relevant for this anomaly 
but cannot exclude it completely. 
In particular, lower limits on $\tau$ obtained from the same experiment \cite{Serebrov2} 
with horizontal and homogeneous magnetic field $B=0.2$~G 
(solid red contour in Fig.~\ref{fig:global}) are compatible with the above anomaly 
for the range $B' = (0.08\div0.35)$~G with respective values of $\tau$ ranging 
from 5~s to 55~s. 
Results of Ref. \cite{Altarev} (blue solid and dashed contours in Fig.~Fig.~\ref{fig:global}) 
yield the lower limits $\tau > 12$~s and $\tau_\beta > 15$~s for any $B'$ less than 0.13~G. 
Combined with the above bounds of Ref. \cite{Serebrov2}  with "horizontal" measurements, 
the limit $\tau_\beta \geq \tau > 12$~s can be extended got the range of mirror fields 
up to $B' = 0.25$~G. 
 
Our experimental results enhance these experimental limits, and also for a wider range of possible 
values of $B'$ (see yellow shaded regions in Fig.~\ref{fig:global}). 
Namely, for any $B'$ in the interval $(0.08 \div 0.17)$~G we get a 
lower limit don $n-n'$ oscillation time $\tau > 17$~s (95 \% C.L.), 
and  $\tau_\beta > 27$~s for any $B'$ in the interval $(0.06 \div 0.25)$~G. 
Assuming that the mirror magnetic field $B'$ is constant in time, 
or in more precise terms, that its value and direction did not change significantly in time 
during the years passed form previous experiments to our measurements,  
we can combine our results with limits of previous works. 
Yet, we could not completely exclude the parameter areas of interest, and 
pink shaded regions in Fig.~\ref{fig:global}) correspond to regions which can still be 
relevant for above mentioned $5\sigma$ and $3\sigma$ anomalies. 
For larger values of $B'$ the limit on $\tau$ and $\tau_\beta$
is considerably weakened and for $B > 0.5$ G the values of $n-n'$ oscillation time  as small as 1 second 
become allowed. 

The following remark is in order.  The results of different experiments performed in different times 
can be combined only if one assumes that mirror magnetic field 
is constant in time. However, this is most naive assumption, which means that the rotations of 
the Earth and the the Baby "mirror Earth" in its interior are completely synchronous, so that 
the orientation of the mirror magnetic field in the given experimental site does not change in time. 
On the other hand,  if the axis of mirror dipole is deviated from the rotation axis, and there is 
some difference between angular velocities by which ordinary and magnetic fields precess, 
then one expect some periodic time variation, then the periodic variation of the signal 
can be expected.  In addition,The  captured mirror matter in the deep interior of the Earth  
can come into thermal equilibrium with the normal matter and thus can be present in the ionized form, 
due to a temperature of the Earth core of several thousand K. 
Also the dynamo mechanism in the differentially rotating mirror plasma could plays a substantial role, 
mirror magnetic field can strongly increase and also its configuration can change from dipole to multipolar or toroidal configuration, with the subsequent inversion of the direction of the mirror dipole. 
 In this the mirror magnetic field of the Earth 
may have also substantial large period time variation, perhaps of few years, like the sun's magnetic field. 
 which can be increased up to several Gauss. 
Unfortunately, time duration of our experiment (one reactor cycle) 
is not enough to place limits on possible long period time variation of mirror magnetic field. 
However, the possibility of time varying mirror field background 
should be taken in consideration while planning the next experiments searching for $n-n'$ 
oscillation, as e.g. $n\to n' \to n$ regeneration experiment with cold neutrons at the stage of preparation 
at HFIR Reactor at Oak Ridge national Laboratory \cite{ORNL}.

\vspace{6mm} 
\noindent{\bf Acknowledgements} 
\vspace{2mm} 

We are grateful to the Institut Laue--Langevin (ILL) for providing the EDM beamline 
on PF2 facility for one reactor cycle  and for excellent technical support throughout the experimental runs. 
The work was supported in part by CETEMPs at the University of L'Aquila.  
Z.B. and R.B. thank the ILL for partial financial support. 
We are grateful to Anatoly Serebrov for help and valuable advices. 
Z.B. thanks Yuri Kamyshkov, Valery Nesvizhevsky  and Guido Visconti for interest to the work 
and useful discussions. 
The results of this work were reported by Z.B. at the Workshop 
INT-17-69W ``Neutron Oscillations: Appearance, Disappearance, and Baryogenesis", 
Institute of Nuclear Theory, Seattle,  USA, October 23 - 27, 2017, 
and at the seminar at Oak Ridge National Laboratory, Knoxville, USA, 31 Oct. 2017. 
We thank Klaus Kirch  for informing us that a new experiment 
on search of neutron -- mirror neutron oscillation was performed at the PSI, Villingen, 
results of which are currently under analysis. 

\noindent


\end{document}